\documentclass[useAMS,usenatbib]{mn2e}
\usepackage{graphicx,fleqn}
\DeclareGraphicsExtensions{.eps,.ps,.eps.gz,.ps.gz,.eps.Z}
\title[ULX in NGC4559]{Probable intermediate mass black holes in NGC4559: \xmm\/ spectral and timing constraints}

\author[]{Mark Cropper$^{1}$, Roberto Soria$^{1}$, Richard F. Mushotzky$^{2}$, Kinwah Wu$^{1}$, \and Craig B. Markwardt$^{3,4}$, Manfred Pakull$^{5}$ \\
$^{1}$ Mullard Space Science Laboratory, University College London,
Holmbury St. Mary, Dorking, Surrey, RH5 6NT, UK\\
$^{2}$ Laboratory for High Energy Astrophysics, Code 660, Goddard Space Flight Center, NASA, Greenbelt, MD, 20 770, USA\\
$^{3}$ Department of Astronomy, University of Maryland, 
	College Park, MD 20742, USA\\
$^{4}$ Laboratory for High Energy Astrophysics, 
        Mail Code 662, NASA Goddard Space Flight Center, Greenbelt, MD 20771, USA\\
$^{5}$ Observatoire Astronomique de Strasbourg, 11 rue de l'Observatoire,
F-67000 Strasbourg, France}

\date{Received: }

\begin{document}
 
\newcommand{\dg} {^{\circ}}
\outer\def\gtae {$\buildrel {\lower3pt\hbox{$>$}} \over
{\lower2pt\hbox{$\sim$}} $}
\outer\def\ltae {$\buildrel {\lower3pt\hbox{$<$}} \over
{\lower2pt\hbox{$\sim$}} $}
\newcommand{\ergscm} {ergs s$^{-1}$ cm$^{-2}$}
\newcommand{\ergss} {ergs s$^{-1}$}
\newcommand{\ergsd} {ergs s$^{-1}$ $d^{2}_{100}$}
\newcommand{\pcmsq} {cm$^{-2}$}
\newcommand{\ros} {{\it ROSAT}}
\newcommand{\xmm} {\mbox{{\it XMM-Newton}}}
\newcommand{\exo} {{\it EXOSAT}}
\newcommand{\sax} {{\it BeppoSAX}}
\newcommand{\chandra} {{\it Chandra}}
\newcommand{\hst} {{\it HST}}
\def\rchi{{${\chi}_{\nu}^{2}$}}
\newcommand{\Msun} {$M_{\odot}$}
\newcommand{\Mwd} {$M_{wd}$}
\newcommand{\Mbh}{$M_{\bullet}$}
\def\Mdot{\hbox{$\dot M$}}
\def\mdot{\hbox{$\dot m$}}
\def\mincir{\raise -2.truept\hbox{\rlap{\hbox{$\sim$}}\raise5.truept
\hbox{$<$}\ }}
\def\magcir{\raise -4.truept\hbox{\rlap{\hbox{$\sim$}}\raise5.truept
\hbox{$>$}\ }}

\maketitle

\begin{abstract}
We have examined X-ray and optical observations of two ultra-luminous X-ray sources, X7 and X10 in NGC4559, using \xmm, \chandra\/ and \hst. The UV/X-ray luminosity of X7 exceeds $2.1\times10^{40}$ erg s$^{-1}$ in the \xmm\/ observation, and that of X10 is $>1.3\times10^{40}$ erg s$^{-1}$. X7 has both thermal and power-law spectral components, The characteristic temperature of the thermal component is $0.12$ keV. The power-law components in the two sources both have slopes with photon index $\simeq 2.1$. A timing analysis of X7 indicates a break frequency at 28 mHz in the power spectrum, while that for X10 is consistent with an unbroken power law. The luminosity of the blackbody component in the X-ray spectrum of X7 and the nature of its time variability provides evidence that this object is an intermediate mass black hole accreting at sub-Eddington rates, but other scenarios which require high advection efficiencies from a hollowed-out disk might be possible. The emission from X10 can be characterised by a single power-law. This source can be interpreted either as an intermediate mass black hole, or as a stellar-mass black hole with relativistically-beamed Comptonised emission. There are four optical counterparts in the error circle of X7. No counterparts are evident in the error circle for X10. 
\end{abstract}


\section{Introduction}
\label{sec:intro}

It has been known for some time that there are point-like, non-nuclear sources
in some nearby galaxies, with luminosities exceeding $10^{39}$ erg s$^{-1}$
(eg. Long, 1982, Fabbiano 1989, Colbert \& Mushotzky 1999, Roberts \& Warwick
2000, Makishima et al. 2000). The nature of these ultra-luminous compact X-ray
sources (ULX) is still unclear.  Fabbiano (1989) showed statistically that they
are not foreground objects, or background AGNs.  Some of the objects have been
identified with young supernova remnants (Fabbiano \& Trinchieri 1987), but many exhibit significant variability on long time scales. Others may be white dwarfs emitting via nuclear burning on their surfaces (Fabbiano et al. 2003). Low
signal-to-noise spectra obtained by {\it Chandra} and {\it ASCA} show at least
two classes of spectra: a pure power-law and a 
thermal blackbody-like spectrum (Strickland et al.\ 2001, 
Makishima et al.\ 2000). They are most likely to be accreting compact
objects (Colbert \& Mushotzky 1999), but because their luminosity exceeds the
Eddington luminosity of a 20 \Msun\/ compact object ($3\times10^{39}$
ergs s$^{-1}$), sometimes by a factor of $\sim15$, this scenario suggests that the compact  object may be 
an intermediate mass black hole (IMBH) of mass \Mbh$\sim100$\Msun.

King et al. (2001) argued that
the production of IMBH is difficult, especially when taking into account the constraints on the mass-donor companion stars. 
However, they can be formed directly from the collapse of high mass stars in low metallicity environments (Heger et al 2003). ULX may also be numerous in elliptical galaxies (Angelini, Loewenstein \& Mushotzky 2001) although this is still unclear (Irwin, Athey \& Bregman 2003).  An extreme case is their creation as a primordial population in the first burst of star formation (Population III stars, Madau \& Rees 2001).  Alternatively, indirect channels may exist for the formation of IMBH, either through binary evolution (Fryer \& Kalogera  2001) or through the mergers of massive stars in Superstar Clusters  (Portegies Zwart \& McMillan 2002). IMBH may also be formed through the mergers of lower mass black hole (BH) in cluster environments (Lee 1995, Miller \& Hamilton 2002).

For higher metallicities, BH masses tend to be \ltae$20$ \Msun. Direct-collapse black holes (the most likely origin of $\gamma$-Ray Bursts) have probable progenitors which yield BH remnant masses with \Mbh$\simeq 14-23$ \Msun\/ (Fryer 1999). If other effects such as disk winds and magnetic fields are important, these masses
could be lower. (For stars with insufficient angular momentum, weak $\gamma$-Ray Burst explosions or no explosions are produced, and the remnant can be much more
massive, up to the mass of the star.) So, if ULX are powered by a stellar mass BH (which we define here as \Mbh$\leq20$\Msun), beaming or super-Eddington luminosities are required to explain the  observed luminosities.  
    
Two different types of beaming can in principle explain the 
luminosity of ULX in external galaxies. The beaming could be due to anisotropic emission of the radiation (King et al.\ 2001), or the relativistic motion of the 
radiation sources (for example Fabrika \& Mescheryakov 2001). 

King et al.\ (2001) pointed out that if the accretion disk around 
the BH has a much lower scattering optical depth over a restricted 
range of solid angle ($\Delta \Omega$) than in other directions, 
almost all the emitted X-rays will emerge in these directions. For 
a beaming factor $b (=\Delta \Omega / 4 \pi) $, the accreting BH 
could have a luminosity $L = b L_{\rm sph}$, where $L_{\rm sph}$ is 
the luminosity inferred in the assumption of spherical accretion/isotropic emission. Such 
beaming is due to the special geometry of the accretion disk, thus  
it does not require the emitting plasma to be in relativistic motion 
relative to the observer. The spectrum of the emission could be 
thermal blackbody-like or a power-law. The chance of discovering  
such sources depends on the opening angle of the beaming cone, the duty cycle and the life span of the phases in which the mass transfer rate from the donor star is high enough to account for the observed X-ray luminosity.

The other type of beaming is caused by the relativistic motion of 
the emitting plasmas. This would be the case, for example, if we were viewing the emission from BL Lac objects or the jets of SS~433-type systems at 
an angle almost parallel to the jet orientation (Fabrika \& Mescheryakov 2001, K\"{o}rding, Falcke, \& Markoff 2002, Georganopoulos, Aharonian \& Kirk 2002). The emission is Doppler-boosted by the 
relativistic motion, and it appears to be ultra bright, 
depending on the jet speed. As the emission is likely to be 
non-thermal direct synchrotron emission or (optically thin) 
Comptonised emission from relativistic beamed electrons, we 
would expect a power-law spectrum instead of a blackbody-like spectrum. The radiation should have strong linear 
polarization (even in the X-ray bands). The discovery 
probability of these type of sources is restricted by the 
opening angle of the jets. Fabrika \& Mescheryakov (2001) estimated 
that about one such relativistic beamed source would be observed 
in each irregular or spiral galaxy, which is similar to the number 
of ULX per galaxy in recent observations. 

It has proved difficult to distinguish whether ULX 
are stellar-mass Galactic BH binaries with beamed X-ray emission 
or the more exotic IMBH systems.  SS~433- and BL Lac-type relativistic beaming models can be excluded for sources with 
a thermal blackbody-like spectrum. Non-relativistic beaming 
(anisotropic emission) models will have difficulties if the X-ray 
observed luminosities are too high. Moreover, these kind of models 
would impose some signatures on the spectra either by Compton 
scattering or photoelectric absorption. As the geometrical beaming model requires 
that the accretion disk is viewed almost face-on, any spectral 
features are expected to indicate a low inclination geometry. 
If the geometrical beaming model is correct, the power spectrum of variability would be shifted to higher frequencies than in the IMBH model because the characteristic frequencies at the inner edge of the disk scale as 1/\Mbh. For the relativistic beaming model, large 
amplitude variability should occur if the emission is direct 
synchrotron radiation. 

\begin{figure*}
\begin{center}
\includegraphics[width=\columnwidth]{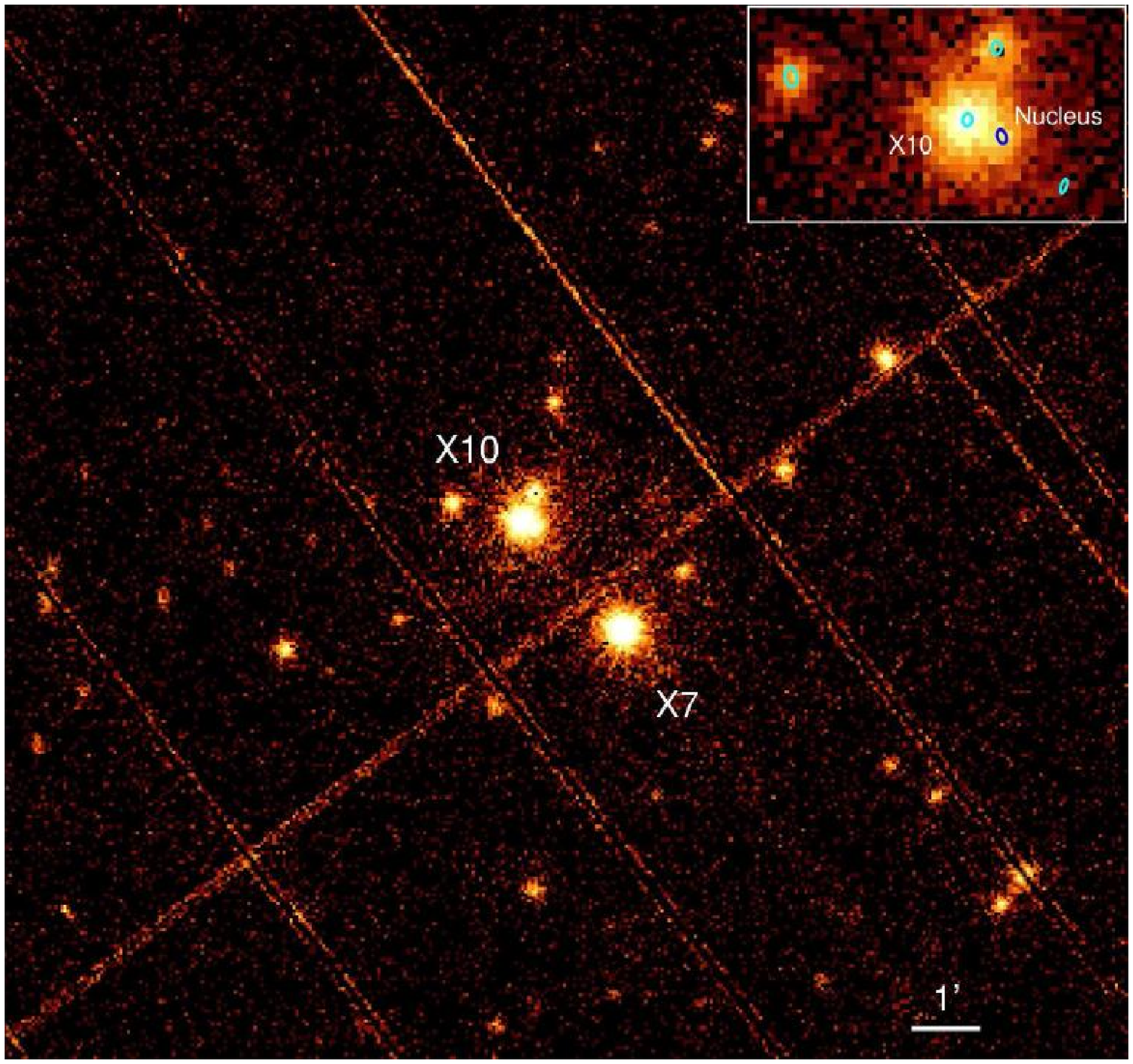}
\hspace*{0.75\columnsep}
\includegraphics[width=\columnwidth]{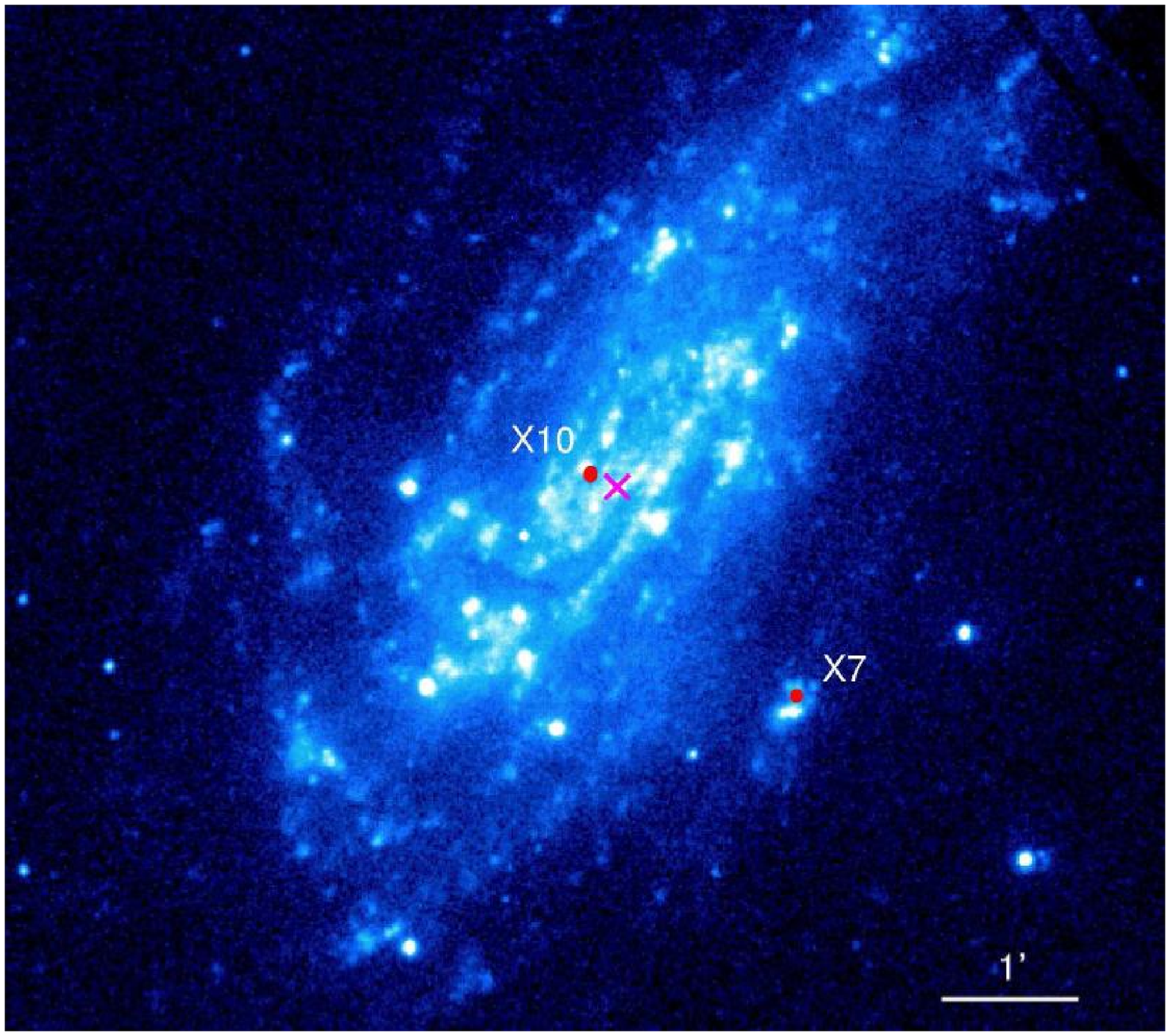}
\end{center}
\caption{(a) The  \xmm\/ EPIC-pn image of NGC4559. North is up and East is left. X7 is the bright source near the centre with the slightly fainter X10 located above and to the left. The inset shows the region around X10 and nucleus, with the \chandra\/ positions superimposed. (b) The \xmm\/ OM UVW1 image of NGC4559. The locations of X7 and X10 are  marked with arrows and small red circles and the centre of the galaxy is marked with a magenta cross.}
\label{fig:xmm_image}
\end{figure*}

Further progress on ULX therefore depends on a detailed spectral study,
a temporal study extending over as wide a range of frequencies as possible and a study of the location of the ULX in the host galaxies, in particular with respect to star forming regions. Ultimately, kinematic studies of confirmed optical counterparts will prove to be definitive.

It almost goes without saying that ULX are of significant interest.  If they
are IMBH, they are the link between stellar mass BH and supermassive BH. Their
origin may be primordial, or the result of stellar or BH mergers. They will have
a major impact on their surroundings. If they are beamed stellar mass BH, the
beaming mechanism and the underlying structure of the disk are also
astrophysically important, and again, the scaling to AGN-generated jets is of
interest.

\section{Observations}

This study concentrates on the study of two ULX in NGC4559, X7 and X10 (Vogler, Pietsch \& Bertoldi 1997). Using the \ros\/ PSPC, Vogler et al. (1997) identified seven X-ray sources falling within the $D_{25}$ ellipse of the galaxy. The two brightest sources, X7, and X10, had intrinsic luminosities $\approx10^{40}$ erg s$^{-1}$ in the 0.1--2.4 keV \ros\/ band. X7 was located in the outer spiral arms, near an H{\sc ii} region. They suggested that X7 could be a young supernova remnant in a dense cloud. X10 was thought to be associated with the nucleus of the galaxy, although this source was noted to be extended, probably due to the superposition of several sources, and offset from the optical nucleus by approximately 15 arcsec.   

Here we follow the Vogler et al. (1997) nomenclature; they are also known as X1 and X4 respectively in Roberts \& Warwick (2000), and IXO65 and IXO66 in Colbert \& Ptak (2002). Roberts \& Warwick (2000) commented that X10 is a nuclear source, while, from careful positional adjustments of \ros\/ HRI observations, Pakull \& Mirioni (2002) recorded it as a ULX displaced from the nucleus. We confirm here that X10 is non-nuclear.

We make use of data obtained with \xmm, \chandra\/ and \hst. The log of observations is given in Table~\ref{tab:log}. The \chandra\/ and \hst\/ data were retrieved from public archives.

\begin{table*}
\begin{center}
\begin{tabular}{lllllll}
\hline
Observation & Observatory 	& Instrument 	& Observation & Duration 	& Filters   & Mode  \\
	Date		&     			& &	ID		& (sec)		&               &   \\
\hline
1994-12-29 & \hst\/ (X10 field)	& WFPC2	& U29R2B01T$\rightarrow$02T	& 160	&  F606W  & \\
2001-01-14 & \chandra	& ACIS-S	& 2026 & 9400	&		&    \\
2001-05-25 & \hst\/ (X7 field)	& WFPC2	& U6EH0505R$\rightarrow$08R & 2000	&  F450W  & \\
		& 			& WFPC2	& U6EH0501R$\rightarrow$04R 	& 2000	&  F555W   & \\
		& 			& WFPC2	& U6EH0509R$\rightarrow$0CR 	& 2000	&  F814W   & \\
2001-06-04 & \chandra 	& ACIS-S 	& 2027	& 10700	& 	 	&   \\
2001-11-13 & \hst\/ (X10 field) & WFPC2	& U6EA7101M$\rightarrow$02M 		& 460	&  F450W  & \\
		& 			& WFPC2		& U6EA7103M$\rightarrow$04M 
		& 460	&  F814W  & \\
2002-03-14 & \chandra 	& ACIS-S 	& 2686	& 3000	& 	 	&   \\
2003-05-27 & \xmm	& EPIC-pn & 0152170501 & 40794 & medium & full-frame \\
	     	& 			& EPIC-MOS1+2 & & 42130 & medium & small-window (X7 only) \\		& 			& OM	        & & 41629   & UVW1 & image+(X7 only)fast  \\
\hline
\end{tabular}
\caption{Log of the observations.}
\label{tab:log}
\end{center}
\end{table*}

\subsection{\xmm\/ data}

The \xmm\/ data were processed using SASv5.4. The \xmm\/ observations were centred on X7. The EPIC-MOS cameras were operated in small window mode to provide sufficient time resolution, with the result that X10 was not seen in EPIC-MOS. The background was generally low during the observation, except at the beginning and for short episodes during the run. These intervals were excised from the data, reducing the effective exposure to 33.9 ksec for EPIC-pn and 36.8 ksec for EPIC-MOS. The image from the EPIC-pn camera is shown in Figure~\ref{fig:xmm_image}a. The 20 UVW1 (effective band 2400--3200\AA) OM exposures each of 2 ksec duration were co-added (Figure~\ref{fig:xmm_image}b).

An extraction circle of  30 arcsec radius was used for X7. For X10, we used a more complex extraction region of 20 arcsec radius with an exclusion of 6 arcsec radius centred on the nucleus of the galaxy 13 arcsec away from X10. This reduced the contamination from the relatively fainter nucleus and another source 28 arcsec to the North and slightly West of X10 \footnote{In fact this source, which we call CXOU J123557.7+275807  is itself sufficiently bright to be classed as a ULX. With an absorbed power-law spectral fit, we estimate the emitted luminosity to be $3 \pm1\times10^{39}$ erg s$^{-1}$ in the 0.3--12 keV band (approximately constant over the \chandra\/ and \xmm\/ observations).
Using an equally good {\sc diskbb} fit, the inferred emitted 
luminosities are lower, $\simeq1.5-2.0\times10^{39}$ erg s$^{-1}$ in the same band. Both this source and the nucleus are not resolved from X10 in Vogler et al. (1997).} to a low level. Background was selected from source-free regions on the same CCD for the EPIC-pn data. For the EPIC-MOS X7 data, the small window mode prevented use of the same CCD for background subtraction, so regions from peripheral CCDs as close as possible to the source were selected. For the extraction, events were filtered according to event patterns 0--4 for EPIC-pn and 0--12 for EPIC-MOS (single and double events in each case), and bad events were rejected using \#XMMEA\_PN=0 and XMMEA\_EM=0. {\it arfgen} and {\it rmfgen} were used to create the response files. We checked that neither source is piled up.

\begin{figure}
\begin{center}
\includegraphics[height=\columnwidth,angle=270]{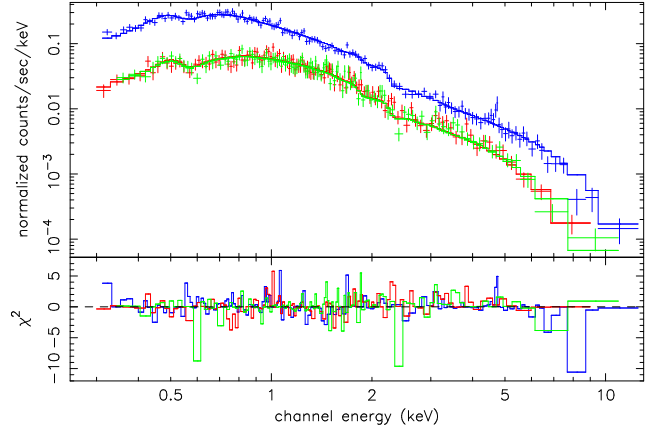}
\end{center}
\caption{The \xmm\/ EPIC-pn (blue, top), EPIC-MOS1(red) and EPIC-MOS2 (green) spectra of X7 simultaneously fitted with the  {\sc wabs(tbvarabs(bb+po))} model in Table~\ref{tab:xmm_fits_x7}.}
\label{fig:xmm_spec_x7}
\end{figure}

The extracted spectra for X7 from the three EPIC cameras are shown in Figure~\ref{fig:xmm_spec_x7}.  We fitted the three datasets simultaneously using XSPECv11.2 (Arnaud 1996).  First we tried a simple power-law model, absorbed according to the {\sc tbvarabs} (T\"{u}bigen-Boulder variable absorption) absorber model with Solar abundance. This does not provide an acceptable fit, $\chi_{\nu}^2=1.8$ Then we reduced the metal abundance to 0.1 of the Anders \& Grevesse (1989) values. This provided a relatively poor fit, $\chi_\nu^2=1.12$. Allowing the abundance to be fitted results in a lower \mbox{$\chi_{\nu}^2=398/381=1.04$} but with abundances  $< 0.01$ of the Anders \& Grevesse (1989) values, which would imply extremely low metallicities for the environment. We have no information on what the metallicities might be, but unless they are extremely low, this indicates that an additional soft component is required. 
Adding a soft component improves the fit significantly, $\chi_{\nu}^2=370/379=0.98$, so that the component is significant at the $> 99.99$ percent level. 
Blackbody plus power-law ({\sc bb+po}), disk blackbody plus power-law ({\sc diskbb+po})  and Comptonised blackbody ({\sc bmc})\footnote{The Comptonisation model {\sc bmc} (Borozdin et al.~1999;
Shrader \& Titarchuk 1999) assumes a blackbody spectrum
for the seed photon component, and is therefore preferable
to the older {\sc comptt} model which uses only the Wien tail
of the blackbody spectrum.} models, all with a {\sc tbvarabs} absorber, all provide equivalently good fits to the data (in each case a {\sc wabs} absorber model was also included to account for Galactic absorption). We found that the normalisation of the fits for the two EPIC-MOS cameras compared to the EPIC-pn is a factor $\sim0.93$ of that expected, so we allowed this relative normalisation to be determined from the fits. The parameters derived from these fits are given in Table~\ref{tab:xmm_fits_x7}. There and elsewhere the units of normalisation in the fitted models are as defined in XSPEC. The low temperature of the thermal component means that we cannot distinguish between single and multi-temperature blackbody models in the EPIC band. The fit for the {\sc bb+po} model is shown running through the data in the top panel of Figure~\ref{fig:xmm_spec_x7}, while lower panel shows the $\chi^2$ residuals from the fits. 

\begin{figure}
\begin{center}
\includegraphics[height=\columnwidth,angle=270]{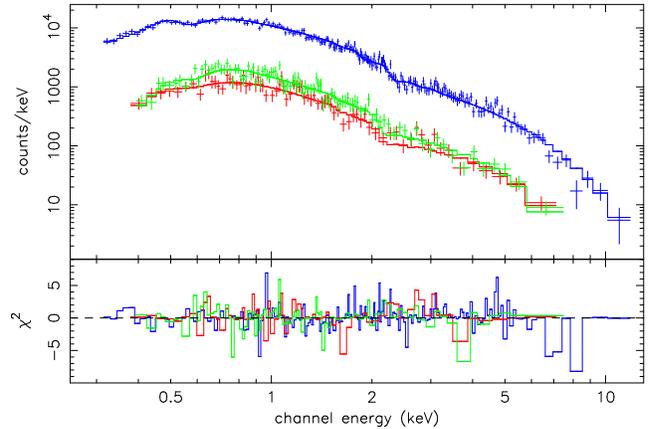}
\end{center}
\caption{The \xmm\/ EPIC combined spectrum of X7 (blue, top) fitted with the  {\sc wabs(tbvarabs(bb+po))} model in Table~\ref{tab:xmm_fits_x7} and \chandra\/1 (bottom, red) and \chandra\/2 (middle, green) spectra fitted individually with the {\sc wabs(tbvarabs(bb+po))} model in Table~\ref{tab:fits_comb_x7}. }
\label{fig:all_spec_x7}
\end{figure}

\begin{table}
\begin{center}
\begin{tabular}{p{0.0cm}p{3.7cm}p{3.4cm}}
\hline
	& Parameter 		& Value 	 \\
\hline
\multicolumn{3}{l}{{\bf EPIC-pn, EPIC-MOS1, EPIC-MOS2}} \\[1mm]
\multicolumn{2}{l}{{\footnotesize WABS(TBVARABS(BB+PO))}} &    $\chi_{\nu}^2=0.98$ (369.6/379 dof) \\[1mm]
&  $n_{\rm H}$(intrinsic) 		& $4.3^{+0.9}_{-1.1}\times10^{21}$ cm$^{-2}$ \\[1mm]
& metal abundance & $0.31^{+0.27}_{-0.20}$ \\[1mm]
& kT(blackbody)  	 & $0.12^{+0.01}_{-0.01}$ keV \\[1mm]
& BB normalisation	& $7.6^{+7.5}_{-4.3}\times10^{-6}$ \\[1mm]
& photon index   & $2.23^{+0.06}_{-0.05}$ \\[1mm]
& PO normalisation	& $2.36^{+0.30}_{-0.17}\times10^{-4}$ \\[1mm]
& MOS/pn normalisation	& $0.93^{+0.05}_{-0.03}$ \\[1mm]
& emitted flux (0.3--12keV) & $1.70\times10^{-12}$ erg cm$^{-2}$ s$^{-1}$ \\
& blackbody flux (0.3--12keV) & $0.43\times10^{-12}$ erg cm$^{-2}$ s$^{-1}$ \\
& blackbody flux (0.01--12keV) & $0.67\times10^{-12}$ erg cm$^{-2}$ s$^{-1}$ \\[1mm]
\multicolumn{2}{l}{{\footnotesize WABS(TBVARABS(DISKBB+PO))}} &    $\chi_{\nu}^2=0.98 $ (372.7/379 dof) \\[1mm]
&  $n_{\rm H}$(intrinsic) 		& $5.1^{+1.4}_{-1.3}\times10^{21}$  cm$^{-2}$ \\[1mm]
& metal abundance & $0.30^{+0.27}_{-0.20}$ \\[1mm]
& kT(inner)  	 & $0.148^{+0.006}_{-0.006}$ keV \\[1mm]
& DISKBB normalisation	& $158^{+340}_{-107}$ \\[1mm]
& photon index   & $2.23^{+0.05}_{-0.04}$ \\[1mm]
& PO normalisation	& $2.37^{+0.30}_{-0.17}\times10^{-4}$ \\[1mm]
& MOS/pn normalisation	& $0.93^{+0.05}_{-0.03}$ \\[1mm]
& emitted flux (0.3--12keV) & $1.97\times10^{-12}$ erg cm$^{-2}$ s$^{-1}$ \\
& diskbb flux (0.3--12keV) & $0.72\times10^{-12}$ erg cm$^{-2}$ s$^{-1}$ \\
& diskbb flux (0.01--12keV) & $1.6\times10^{-12}$ erg cm$^{-2}$ s$^{-1}$ \\[1mm]
\multicolumn{2}{l}{{\footnotesize WABS(TBVARABS(BMC))}} &    $\chi_{\nu}^2=0.98$ (371.9/379 dof)\\[1mm]
&  $n_{\rm H}$(intrinsic) 		& $4.4^{+0.7}_{-0.7}\times10^{21}$ cm$^{-2}$ \\[1mm]
& metal abundance & $0.16^{+0.27}_{-0.20}$ \\[1mm]
& kT(blackbody)  	 & $0.12^{+0.01}_{-0.01}$ keV \\[1mm]
& photon index   & $2.21^{+0.05}_{-0.05}$ \\[1mm]
& log(A)	& $0.06^{+0.08}_{-0.04}$ \\[1mm]
& BMC normalisation	& $1.04^{+0.42}_{-0.20}\times10^{-5}$ \\[1mm]
& MOS/pn normalisation	& $0.93^{+0.05}_{-0.03}$ \\[1mm]
& emitted flux (0.3--12keV) & $1.35\times10^{-12}$ erg cm$^{-2}$ s$^{-1}$ \\
& flux (0.01--12keV) & $1.55\times10^{-12}$ erg cm$^{-2}$ s$^{-1}$ \\
\hline
\end{tabular}
\caption{Parameters for the model fits to the \xmm\/ X7 data. The $n_{\rm H}$ Galactic absorption for the WABS model is fixed at $1.5\times10^{20}$ from Dickey \& Lockman (1990). The metal abundance values are given relative to Anders \& Grevesse (1989) for the absorbing gas in the line of sight within NGC4559. All confidence intervals are 90 percent.}
\label{tab:xmm_fits_x7}
\end{center}
\end{table}

The data from the three EPIC cameras were then co-added according to the prescription in Page, Davis \& Salvi (2003). This spectrum with the fit from the parameters in Table~\ref{tab:xmm_fits_x7}  is shown in Figure~\ref{fig:all_spec_x7}.

In the case of X10, an absorbed power-law model provides an acceptable fit to the data and there is no need for additional components. The derived parameters are given in Table~\ref{tab:xmm_fits_x10}, and the spectrum is in Figure~\ref{fig:all_spec_x10}.

\begin{table}
\begin{center}
\begin{tabular}{p{0.0cm}p{3.7cm}p{3.4cm}}
\hline
			& Parameter 		& Value 	 \\
\hline
\multicolumn{3}{l}{{\bf EPIC-pn}} \\[1mm]
\multicolumn{2}{l}{{\footnotesize WABS(TBVARABS(PO))}} &    $\chi_{\nu}^2=0.85$ (155.9/184 dof)\\[1mm]
                                      &  $n_{\rm H}$(intrinsic) 	& $2.7^{+1.3}_{-1.3}\times10^{21}$ cm$^{-2}$  \\[1mm]
				& metal abundance & $0.33^{+0.53}_{-0.19}$ \\[1mm]
                                      & photon index   & $2.05^{+0.07}_{-0.07}$ \\[1mm]
				& PO normalisation	& $1.97^{+0.16}_{-0.14}\times10^{-4}$ \\[1mm]
				& emitted flux (0.3--12keV) & $1.13\times10^{-12}$ erg cm$^{-2}$ s$^{-1}$ \\
\hline
\end{tabular}
\caption{Parameters for the model fits to the \xmm\/ X10 data. The $n_{\rm H}$ Galactic absorption for the WABS model is fixed at $1.5\times10^{20}$ from Dickey \& Lockman (1990). The metal abundance values are given relative to Anders \& Grevesse (1989) for the absorbing gas in the line of sight within NGC4559.   All confidence intervals are 90 percent.}
\label{tab:xmm_fits_x10}
\end{center}
\end{table}

\subsection{\chandra\/ data}

In addition to the 2003 \xmm\ observations,
NGC4559 was observed three times by {\it Chandra} 
ACIS-S between 2001 January and 2002 March (Table~\ref{tab:log}). 
On all occasions, the back-illuminated S3 chip
was used. We obtained the {\it Chandra} dataset
from the public archive and processed it with
the {\footnotesize{CIAO}} software version 2.3.
We inspected the background count rates during
the exposures, and chose to retain all three intervals
in full. We applied the ACIS CTI correction thread, which also applied the newest 
available gain map.

X7 and X10 are easily detected in all three exposures. Using the {\footnotesize CIAO} {\it wavdetect} task, we obtained image centroids and therefore X-ray positions. In addition, the nucleus of NGC4559 is evident in the first and third \chandra\/ exposures, but is only at most marginally detected in the second. This corresponds to a reduction in emission by a factor $>20$ and a return to the previous level over periods of several months. We used the relative positions of X10 and the nucleus to produce the extraction centroids for the \xmm\/ reduction above. We provide the X-ray positions of X7, X10, the nucleus and the third brightest point source (CXOU~J123557.7+275807) from this analysis in Table~\ref{tab:pos}. The estimated uncertainty is 0.6 arcsec (90 percent confidence), derived from an addition in quadrature of the centroid uncertainty, together with the systematic uncertainty of the spacecraft pointing (see http://cxc.harvard.edu/cal/ASPECT/celmon/). This is confirmed by the coincidence between another \chandra\/ point-source with 
a catalogued, isolated optical counterpart 
which has a clear counterpart
in the USNO-B1.0 Catalog (Monet et al. 2003) 
within 0.4 arcsec (this source may be slightly extended).
Another Chandra source is very close 
to the chip edge, so its astrometry may be affected, but even for this source the USNO-B1.0 Catalogue coincides to 0.16 arcsec.
A discussion of the X-ray sources other than X7 and X10 in this galaxy is left to further work.

\begin{table}
\begin{center}
\begin{tabular}{lll}
\hline
Source & RA 	& Dec  \\
\hline
X7   			& $12^h 35^m 51\fs71 $ & $+27\degr 56^\prime 04\farcs1$  \\
X10  		&  $12^h 35^m 58\fs55$ & $+27\degr 57^\prime 41\farcs8$ \\
nucleus   		& $12^h 35^m 57\fs64$ & $+27\degr 57^\prime 35\farcs8$  \\
CXOU J123557.7+275807 & $12^h 35^m 57\fs77$ & $+27\degr 58^\prime 07\farcs 4$  \\
\hline
\end{tabular}
\caption{Positions for the three brightest point sources and the nuclear source in NGC4559 derived from the \chandra\/ analysis. The estimated uncertainty is 0.6 arcsec (90 percent confidence.}
\label{tab:pos}
\end{center}
\end{table}

We used an extraction radius of 3.5 arcsec, with background regions described by an annulus with inner and outer radii of 5 and 15 arcsec. We used {\it psextract} for extracting the source and background, and building the response and auxiliary response matrices. We also used the {\it acisabs} script provided by the {\it CXC} to correct the auxiliary response function for the time-dependent decreased sensitivity of the ACIS-S detector at lower energies.

Spectral fitting was possible for the first two \chandra\/ observations. The third \chandra\/ observation is too short for a spectral
analysis, so we have used  the countrates to provide only an additional point for the 
long-term intensity variations. 

The two background-subtracted \chandra\/ spectra for X7 are shown in Figure~\ref{fig:all_spec_x7} below the combined \xmm\/ EPIC spectrum. Based on our experience with the \xmm\/ data, we have restricted our model fits to a {\sc bb+po} model, with the {\sc tbvarabs} absorber model. The results are given  in Table~\ref{tab:fits_comb_x7}. 

\begin{table}
\begin{center}
\begin{tabular}{p{0.0cm}p{3.7cm}p{3.4cm}}
\hline
			& Parameter 		& Value 	 \\
\hline
\multicolumn{3}{l}{{\bf \chandra\/ 1}} \\[1mm]
\multicolumn{2}{l}{{\footnotesize WABS(TBVARABS(BB+PO))}} &    $\chi_{\nu}^2=1.16$ (83.5/72 dof)\\[1mm]
&  $n_{\rm H}$(intrinsic) 		& $3.6^{+0.9}_{-1.1}\times10^{21}$  cm$^{-2}$ \\[1mm]
& metal abundance & $0.31$ (from \xmm) \\[1mm]
& kT(blackbody)  	 & $0.12^{+0.06}_{-0.06}$ keV \\[1mm]
& BB normalisation	& $8.5^{+1.0}_{-4.3}\times10^{-6}$ \\[1mm]
& photon index   & $1.80^{+0.08}_{-0.08}$ \\[1mm]
& PO normalisation	& $1.78^{+0.20}_{-0.17}\times10^{-4}$ \\[1mm]
& emitted flux (0.3--10keV) & $1.66\times10^{-12}$ erg cm$^{-2}$ s$^{-1}$ \\
& blackbody flux (0.3--10keV) & $0.52\times10^{-12}$ erg cm$^{-2}$ s$^{-1}$ \\
& blackbody flux (0.01--10keV) & $0.72\times10^{-12}$ erg cm$^{-2}$ s$^{-1}$ \\[1mm]
\multicolumn{3}{l}{{\bf \chandra\/ 2}} \\[1mm]
\multicolumn{2}{l}{{\footnotesize WABS(TBVARABS(BB+PO))}} &    $\chi_{\nu}^2=1.11$ (105.8/95 dof)\\[1mm]
&  $n_{\rm H}$(intrinsic) 		& $5.7^{+0.9}_{-1.1}\times10^{21}$  cm$^{-2}$ \\[1mm]
& metal abundance & $0.31$ (from \xmm) \\[1mm]
& kT(blackbody)  	 & $0.12^{+0.01}_{-0.01}$ keV \\[1mm]
& BB normalisation	& $2.28^{+0.5}_{-0.5}\times10^{-5}$ \\[1mm]
& photon index   & $2.13^{+0.08}_{-0.08}$ \\[1mm]
& PO normalisation	& $2.87^{+0.20}_{-0.17}\times10^{-4}$ \\[1mm]
& emitted flux (0.3--10keV) & $2.82\times10^{-12}$ erg cm$^{-2}$ s$^{-1}$ \\
& blackbody flux (0.3--10keV) & $1.30\times10^{-12}$ erg cm$^{-2}$ s$^{-1}$ \\
& blackbody flux (0.01--10keV) & $1.91\times10^{-12}$ erg cm$^{-2}$ s$^{-1}$ \\
\hline
\end{tabular}
\caption{Parameters for the model fits to the \chandra\/ X7 data. The emitted fluxes are estimated in the 0.3--10 keV  \chandra\/ band. The $n_{\rm H}$ Galactic absorption for the {\sc wabs} model is fixed at $1.5\times10^{20}$ from Dickey \& Lockman (1990) and the metal abundance values are relative to Anders \& Grevesse (1989) and fixed using the \xmm\/ value in Table~\ref{tab:xmm_fits_x7}.  All confidence intervals are 90 percent.}
\label{tab:fits_comb_x7}
\end{center}
\end{table}

\begin{figure}
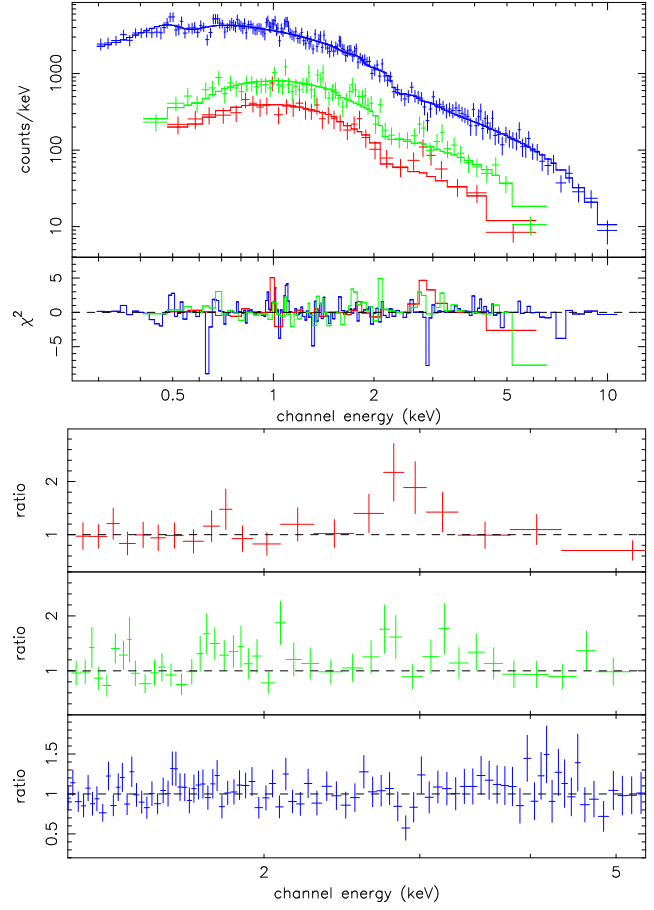

\begin{center}
\includegraphics[height=\columnwidth,angle=270]{cropper_fig4a.ps}
\vspace*{1mm}
\includegraphics[height=\columnwidth,angle=-90]{cropper_fig4b.ps}
\end{center}
\caption{The upper plot shows the \xmm\/ (top, blue),  \chandra\/1 (bottom, red) and \chandra\/2 (middle, green) spectra for X10 together with the spectral fit and residuals using the {\sc tbvarabs(po)} model in Table~\ref{tab:xmm_fits_x10}. Note the residuals around 2.9 keV. These are more clearly evident in the lower plot, which shows the residuals  of the spectrum of X10 from the model fit  in the region around 2.9 keV for the \chandra\/1, \chandra\/2\/ and \xmm\/ observations from top to bottom, respectively. }
\label{fig:all_spec_x10}
\end{figure}

The two background-subtracted \chandra\/ spectra for X10 are shown in Figure~\ref{fig:all_spec_x10} below the \xmm\/ EPIC-pn spectrum. As for the \xmm\/ data, we have restricted our model fits to a {\sc po} model, with a {\sc tbvarabs} absorber model.  The results are given  in Table~\ref{tab:fits_comb_x10}, and the three spectra and fits are shown in Figure~\ref{fig:all_spec_x10}. 

\begin{table}
\begin{center}
\begin{tabular}{p{0.0cm}p{3.7cm}p{3.4cm}}
\hline
		& Parameter 		& Value 	 \\
\hline
\multicolumn{3}{l}{{\bf \chandra\/ 1}} \\[1mm]
\multicolumn{2}{l}{{\footnotesize WABS(TBVARABS(PO))}} &    $\chi_{\nu}^2=0.86$ (27.4/32 dof)\\[1mm]
&  $n_{\rm H}$(intrinsic) 		& $5.6^{+1.8}_{-1.6}\times10^{21}$ cm$^{-2}$  \\[1mm]
& metal abundance & $0.33$ (from \xmm) \\[1mm]
& photon index   & $1.99^{+0.22}_{-0.21}$ \\[1mm]
& PO normalisation	& $1.99^{+0.27}_{-0.21}\times10^{-4}$ \\[1mm]
& emitted flux (0.3--10keV) & $6.6\times10^{-13}$ erg cm$^{-2}$ s$^{-1}$ \\[1mm]
\multicolumn{3}{l}{{\bf \chandra\/ 2}} \\[1mm]
\multicolumn{2}{l}{{\footnotesize WABS(TBVARABS(PO))}} &    $\chi_{\nu}^2=1.08$ (71.2/66 dof)\\[1mm]
&  $n_{\rm H}$(intrinsic) 		& $3.8^{+1.0}_{-0.8}\times10^{21}$  cm$^{-2}$ \\[1mm]
& metal abundance & $0.33$ (from \xmm) \\[1mm]
& photon index   & $1.82^{+0.12}_{-0.13}$ \\[1mm]
& PO normalisation	& $2.03^{+0.26}_{-0.23}\times10^{-4}$ \\[1mm]
& emitted flux (0.3--10keV) & $1.28\times10^{-12}$ erg cm$^{-2}$ s$^{-1}$ \\[1mm]
\hline
\end{tabular}
\caption{Parameters for the model fits to the \chandra\/ X10 data. Emitted flux is  calculated in the 0.3-10 keV band. As for X7, the line of sight Galactic absorption is fixed at $1.5\times10^{20}$ from Dickey \& Lockman (1990).  All confidence intervals are 90 percent.}
\label{tab:fits_comb_x10}
\end{center}
\end{table}

\subsection{Variability}

\subsubsection{Long-term variability}

\begin{figure}
\begin{center}
\includegraphics[height=\columnwidth,angle=270]{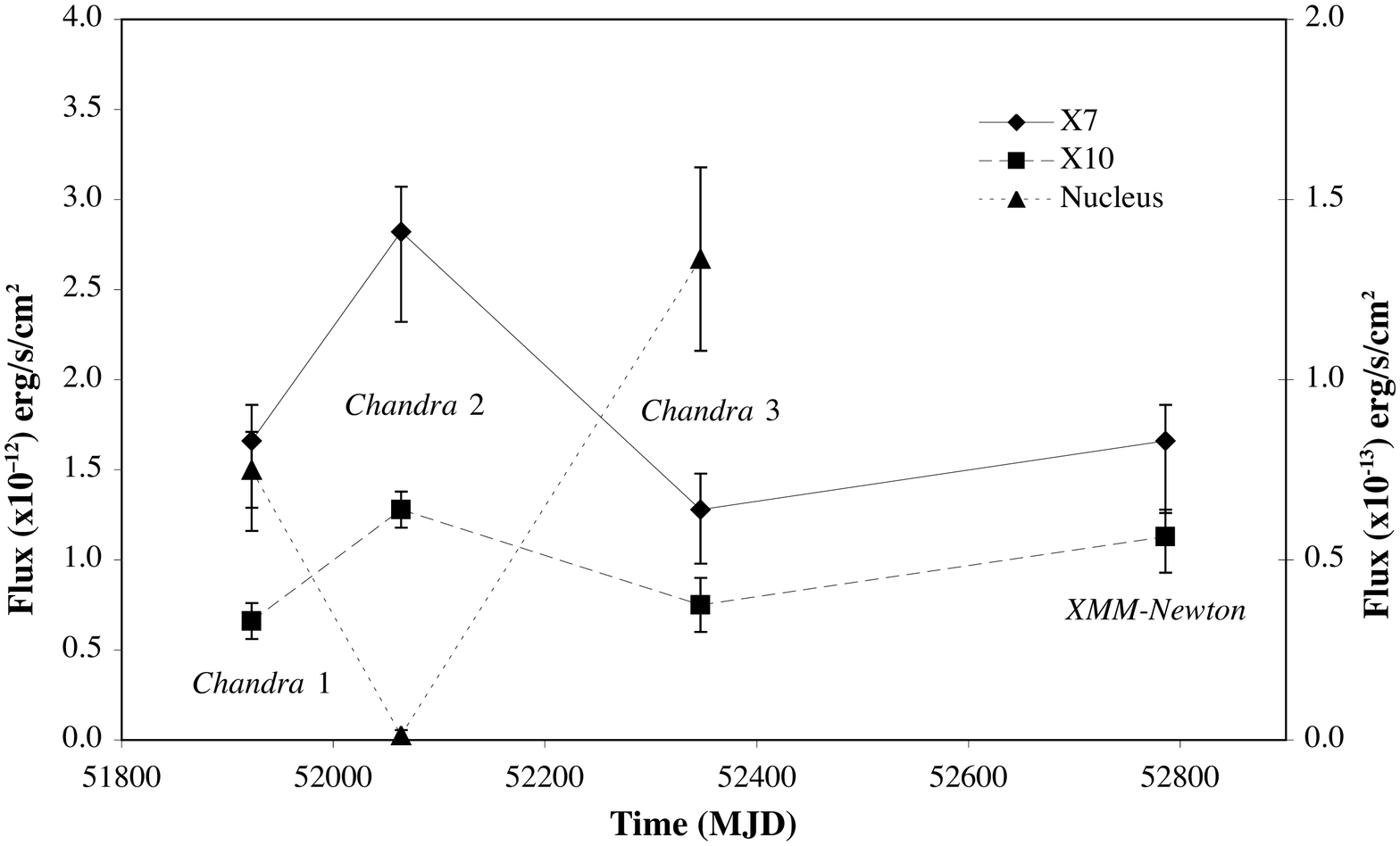}
\end{center}
\caption{The long-term light curves of X7, X10 (left axis) and the nucleus (right axis). The ULX fluxes are determined using the {\sc tbvarabs(bb+po)} and {\sc tbvarabs(po)} fits respectively in the 0.3-10 keV band. The nuclear light curves assume a constant power-law model with photon index 2.2 and $n_{\rm H}$$=3\times10^{21}$. No \xmm\/ point is available for the nucleus because of the contamination from X10 (Figure~\ref{fig:xmm_image}). In terms of counts the nucleus is a factor of $\simeq10$ fainter than X10, which is similar to the level seen in the \chandra\/1 and 3 observations. The error bars indicate 90 percent confidence ranges for the fluxes.}
\label{fig:lt_lc}
\end{figure}

From the fluxes derived from the best fit spectral parameters in Table~\ref{tab:fits_comb_x7} and \ref{tab:fits_comb_x10}, we obtained a long-term light curve for the 0.3--10 keV band for both X7 and X10. In the case of the third \chandra\/ observation, we have derived the fluxes from the count rates assuming the average absorption, temperature and photon index parameters for the first two \chandra\/ observations in Table~\ref{tab:fits_comb_x7} and \ref{tab:fits_comb_x10}. This adds a fourth point. The light curve is shown in Figure~\ref{fig:lt_lc}. 

We note that there is a significant change in the spectral index of the power-law component in X7 over the three observations, while the temperature of the thermal component remains constant. The spectral index of the power-law in X10 remains constant.

The nuclear source is always too faint for spectral fitting in individual exposures. We have analysed the co-added spectrum, which can be fitted by an absorbed power-law with photon index $\simeq 2.2$ and $n_{\rm H}$ $\simeq3\times10^{21}$. We have used these average parameters to convert from count rates to $0.3-10$ keV fluxes. 

\subsubsection{Variability on shorter timescales}
\label{sec:obs_variability}

\begin{figure}
\begin{center}
\includegraphics[height=\columnwidth,angle=90]{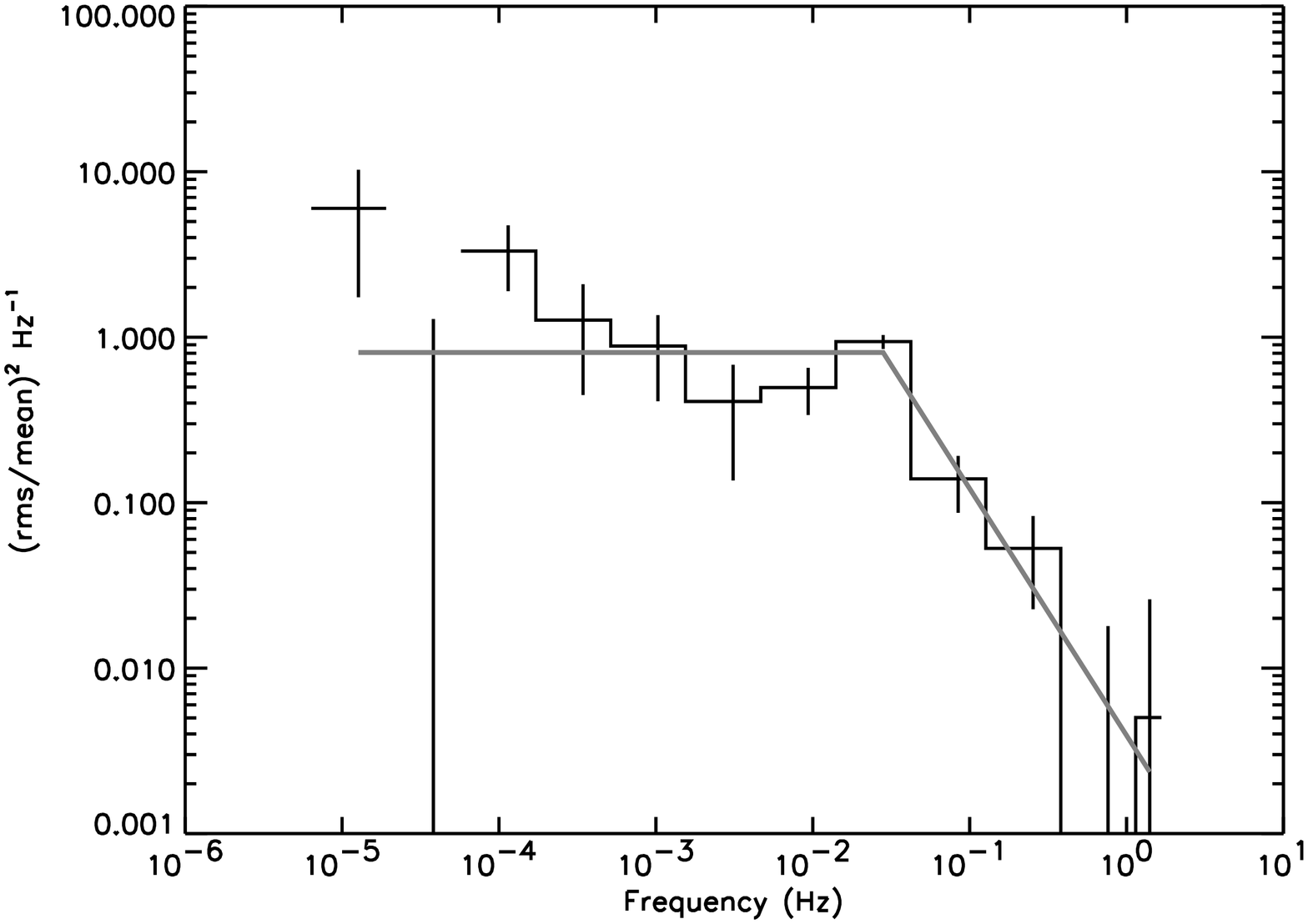}
\includegraphics[height=\columnwidth,angle=90]{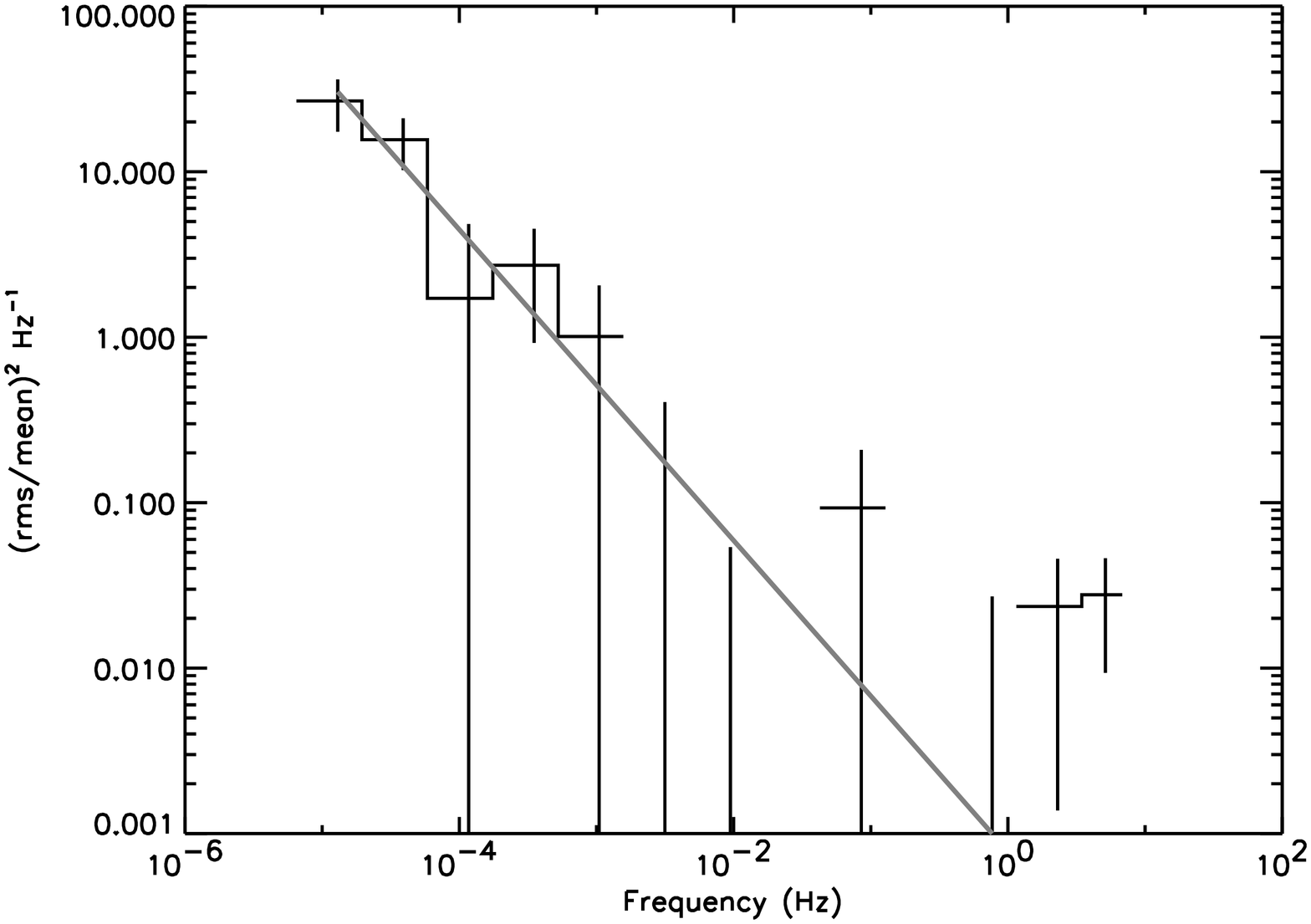}
\end{center}
\caption{The power density spectra for X7 (top) and X10 (below) calculated from the \xmm\/ data.The best fitting model
is shown with a shaded line.  The Poisson noise level has been
subtracted assuming that there is no intrinsic source signal above
a frequency of 1 Hz. }
\label{fig:xmm_pds}
\end{figure}

We examined the variability of X7 and X10 within the \xmm\/ observations using Fourier spectroscopy.
For X10, EPIC-pn events within the energy range 0.3--9 keV (to maximise the signal-to-noise ratio) and within 30 arcsec of
the source were selected, excluding circular regions around the two nearby
sources with radii 10.5 and 8.2 arcsec.  The events were binned into a
light curve with a sampling period of 73.4 ms, the time resolution of
the EPIC-pn detector.  The presence of gaps in a light curve can contribute
significant contaminating power over a wide range of frequencies, and
therefore must be handled properly.  We accounted for the gaps by subtracting the mean count rate (0.21 ct s$^{-1}$ for X10) from the
portions of the light curve with valid data, and assigning a rate of
zero otherwise, which results in a zero-mean light curve. The light curve was zero-padded out to the next highest power of 2. The Fourier
transform was then computed and the squared coefficients were rebinned.
The resulting power spectrum was normalized to be the fractional
mean squared variability per unit frequency interval.  The result is
shown in Figure~\ref{fig:xmm_pds}. 

The spectrum appears to contain some variability at very low
frequencies.  To investigate further, we compared the results of
fitting both a constant and a power-law model ($P(\nu) \propto
\nu^{-\alpha}$) to the spectrum.  The $\chi^2$ for the power-law model
was 10.5 for 11 degrees of freedom, compared to a $\chi^2$ of 25.9
with 12 degrees of freedom for the constant model.  Using an F-test,
we conclude that the variability is significant at the 99.8 percent level.
The best fitted power-law index was $\alpha = 0.9 \pm 0.2$, with a total
r.m.s. variability of $19\pm 7$ percent.

To see if overall background fluctuations could have contributed to
the detected variability (such as low level flaring), we examined a
2800 arcsec$^2$ region of the same EPIC-pn CCD without bright sources.
The mean background rate was 0.057 ct s$^{-1}$, or an expected background
rate of 0.004 ct s$^{-1}$ in the X10 selection region.  Since this rate is
$\sim2$ percent of the X10 mean rate, fluctuations of the background can be
neglected, and the detected variability appears to be real.

To investigate the variability in X7, we performed essentially the
same procedure as for X10, with a few exceptions.  The EPIC-MOS1, -MOS2
and EPIC-pn events were selected within 30 arcsec of the source, and
combined into one joint light curve, with a time sampling of 0.3
seconds (the time resolution of the Small Window image mode used
for the EPIC-MOS observations).  The mean counting rates were 0.29 ct s$^{-1}$ for
the EPIC-pn, and 0.17 ct s$^{-1}$ for EPIC-MOS1 and EPIC-MOS2 combined.  The resulting power
density spectrum is also shown in Figure~\ref{fig:xmm_pds}.  We searched for apparent beats
between the EPIC-pn and EPIC-MOS time sampling frequencies, but found none.

Excess power is clearly detected, with an r.m.s. amplitude of $37\pm 2$ percent.  The improved detection significance is aided by the fact
that the total count rate of X7 is almost a factor of two higher, and
that the intrinsic variability itself is greater.  There appears to be a
break in the spectrum from flat to declining around 30 mHz.   The
precision is greatest at the break as a consequence of
the binning and the change in slope (there are more Nyquist frequency bins per logarithmic bin as the frequency increases, so in general the statistical precision will increase as
long as the signal is constant; however, above the break, the signal
decreases, so the statistical precision decreases).

To treat
this phenomenon more formally, we fitted both unbroken and broken power
law models to the spectrum.  We assumed a flat spectrum ($\alpha_1 = 0$)
below the break frequency, $f_b$.  The unbroken power-law produced a
$\chi^2$ of 81 for 10 degrees of freedom, which improved to 14 for 9
degrees of freedom when the broken power-law model was fitted, a
significant improvement at the 99.99 percent level.  Thus, we conclude
that the break in the power spectrum is a real feature.  The power-law
index above the break was $\alpha_2 = 1.5 \pm 0.8$.  

To test the flatness of the power density spectrum below the break, we performed an additional fit where the lower power-law index was allowed to vary. The best fit power-law index below the break was  $\alpha_1 = 0.20 \pm 0.14$.  The $\chi^2$ value improves to 12.2,  which is not significant enough to rule out an improvement purely by chance.  Neither the upper power-law index, $\alpha_2$, nor the break frequency, $f_b$, changed significantly with the new fit.

Again, we investigated the fluctuations in the background and found
the background rate to be of order 2 percent of the source rate, and
negligible.  In addition, those background fluctuations that were
marginally detectable were very low frequency ($\sim 10^{-5}$ Hz),
very far from $f_b$.  Finally, we examined the power spectrum for the
EPIC-pn and EPIC-MOS data individually, and found that similar variability was
detectable in each, although at reduced significance.  Thus, we are
confident that the break is intrinsic to X7 and not the {\it
XMM-Newton\/} detector systems.

The best fitted break frequency was $f_b = 28$ mHz.  We determined
confidence limits on $f_b$ by computing $\chi^2$ on a grid of $f_b$
values, and allowing the other parameters to vary.  The $1\sigma$
confidence region for $f_b$ is 26--40 mHz.

\subsection{\hst\/ images}

The \hst\/ observations were retrieved from the public archive at ESO produced by the WFPC2  Associations Science Produces Pipeline (WASPP) (see http://archive.stsci.edu/hst/wfpc2/). Data for both the X7 and X10 fields are available in three optical filters (Table~\ref{tab:log}). For X7 the images from each filter were combined to produce a single colour image, while for X10, we have used only the two later exposures. These images are shown in the upper panels of Figure~7. 

We found that the coordinate systems within the WFPC2 files were inconsistent from filter to filter by up to 2 arcsec. We have therefore identified bright stars within the four WFPC2 CCDs corresponding to entries in the USNO A-2 and STSci Guide Star catalogs to recalculate the astrometric solution. Only a small number of stars were available in each case, limiting the correction to a translation. Nevertheless by this means the images can be registered to within 0.3 arcsec.

This allowed us to compare the \chandra\/ and \hst\/ images. Details of the images in the vicinity of the \chandra\/ positions are shown in the lower left panels of Figure~7. Also shown is the error circle calculated by adding in quadrature the positional uncertainty from the \hst\/ images with that from the  \chandra\/ observations (which dominates). This procedure makes clear the environments of  the two ULX. For X7, the 4 brightest stars in the error circle 
have 23  \ltae\ $V$ \ltae\ 24 and 23 \ltae\ $R$ \ltae\ 24.  There are no stars detected 
in the X10 error circle with $R$ \ltae\ 23.5 and $B$ \ltae\ 24.5. More details of this analysis will be presented in Soria et al. (in prep).

The field of X7 was also observed at much lower spatial resolution by Roberts et al. (2002) using the {\it Integral} imager on the William Herschel Telescope. Their images are inverted (North is at the bottom) with respect to ours. We also note that there is a small displacement of $\sim0.7$ arcsec between their position determination and our error circle. Mirioni (2003) shows H$\alpha$ images of both the X7 and X10 fields.

\begin{figure*}
\begin{center}
\includegraphics[width=\columnwidth]{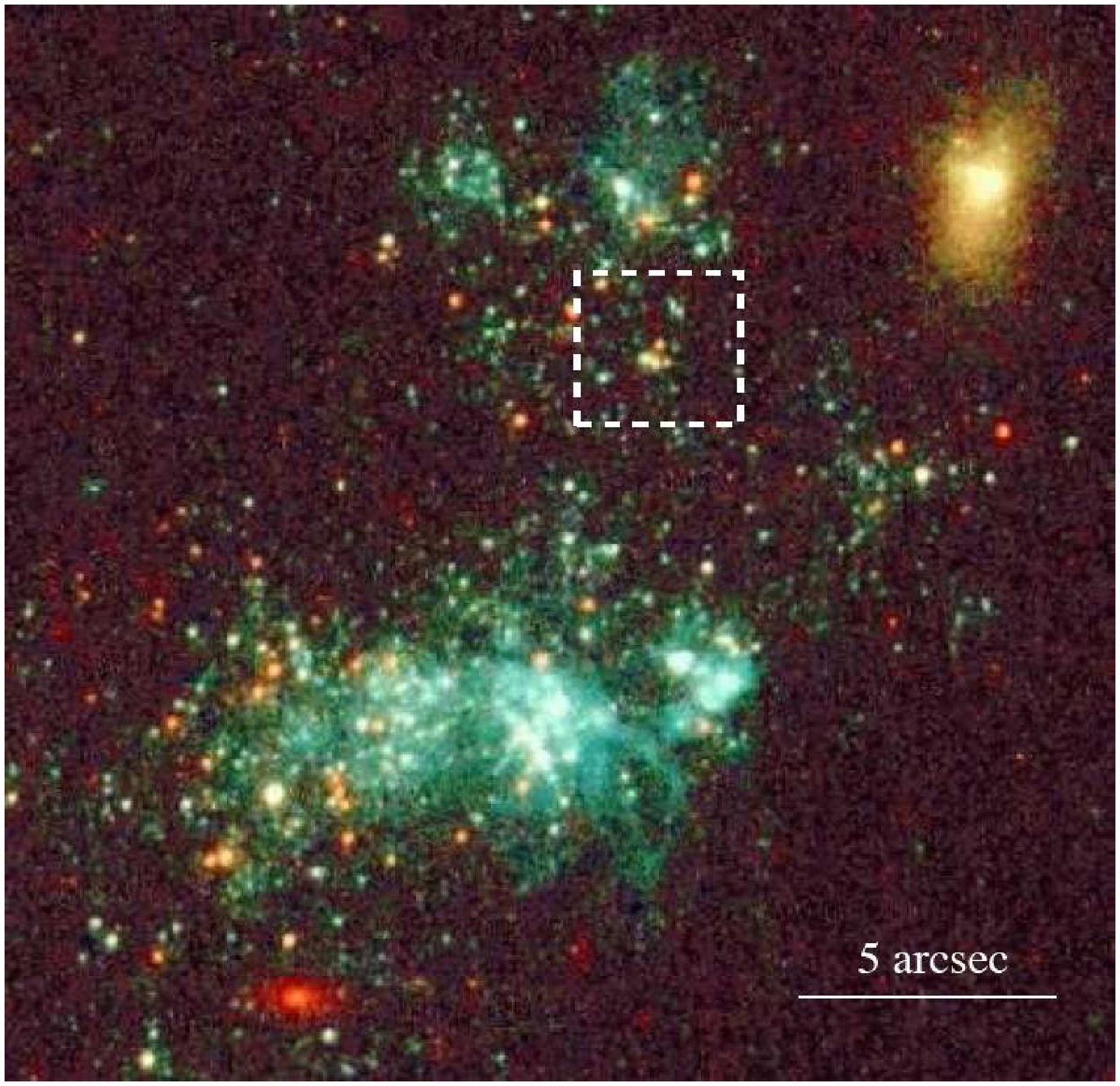}
\hspace{0.5\columnsep}
\includegraphics[width=0.972\columnwidth]{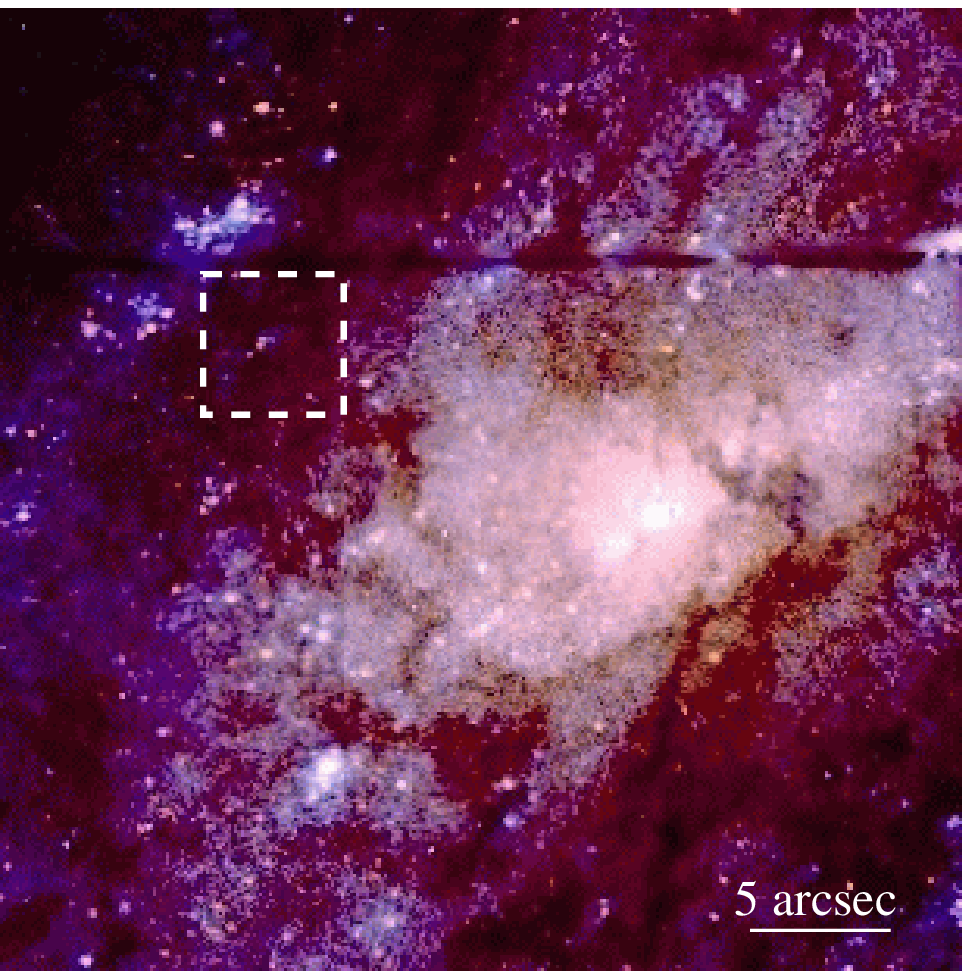}\\
\hspace*{2.4mm}
\includegraphics[width=0.5\columnwidth]{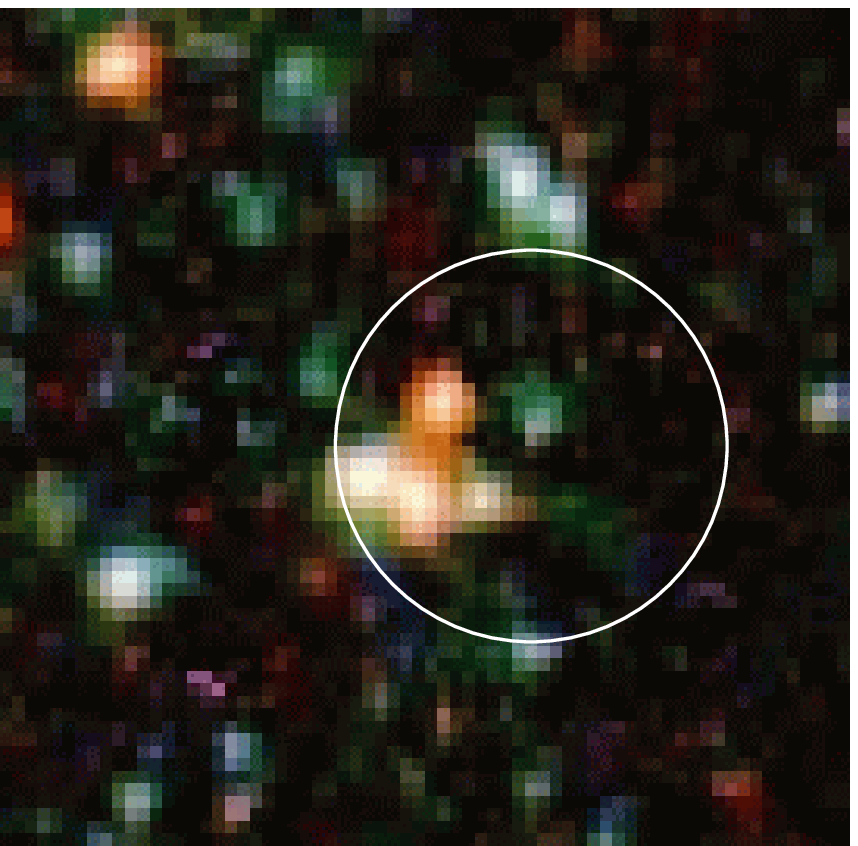}
\hspace{0.55\columnsep}
\includegraphics[width=0.52\columnwidth]{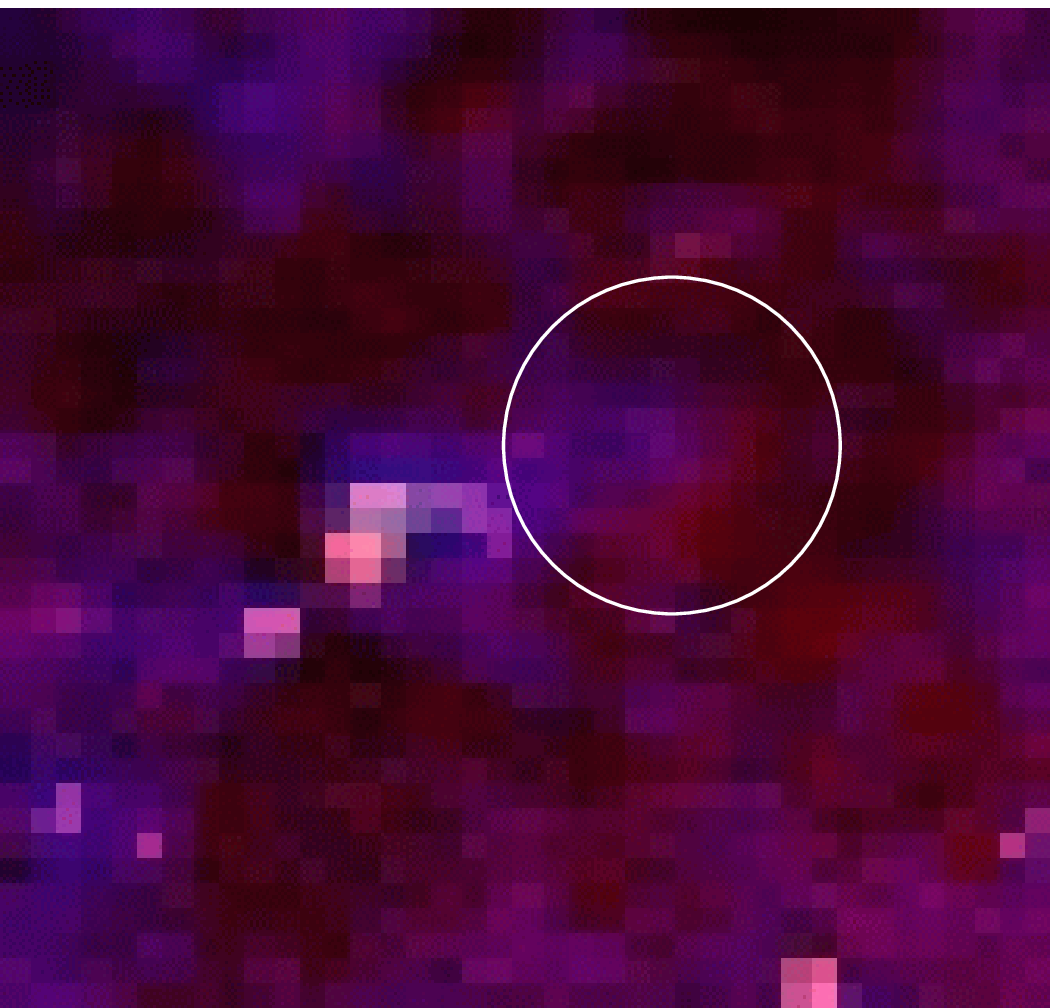}
\caption{HST images of the environments of X7 (3-colour) and X10 (2-colour) with insets showing the 90 percent error circles which are 0.7 arcsec radius.}
\end{center}
\label{fig:hst_colour}
\end{figure*}

\section{Discussion}
\label{sect:disc}

\subsection{Long-term light curves}

Fabbiano et al.\ (2003) argued against an IMBH interpretation 
  for the source CXOAnt J120151.6$-$185231.9, 
  which has similar luminosity to X7, 
  on the grounds that the wide variation in luminosity precluded it (assuming a constant temperature). 
Neither X7 nor X10 show a range more than a factor two in our data (Figure~\ref{fig:lt_lc}),  
  but our sampling is of course limited. Roberts et al. (2002) show the X7 light curve over a longer period including \ros\/ points, together with some \chandra\/ points. We have retrieved the \ros\/ data from the archive, but find that the relative scaling between these and the \chandra\/ and \xmm\/ points is uncertain by a factor of $\sim2$ because of the smaller energy range in \ros,  the lack of spectral information in the \ros\/ HRI data and the confusion with CXOU~J123557.7+275807. We have therefore not incorporated them. Nevertheless, between Figure~\ref{fig:lt_lc} and Roberts et al. (2002), the long-term variability of X7 is evidently modest.

As noted in Section~\ref{sec:intro}, 
  one possibility for the nature of ULX
  is that they are supernova remnants. 
Such variability as is observed in (Figure~\ref{fig:lt_lc}), and its non-monotonic nature,   excludes the possibility that X7 and X10 are supernova remnants.

\subsection{Luminosity}

Assuming a distance of 9.69~Mpc (Sanders et al.\ 2003) 
  and a Solar abundance, 
  we obtain from the fluxes measured 
  in the \xmm\/ data in (Table~\ref{tab:fits_comb_x7}) 
  that the isotropic luminosity of X7 in the 0.3--12keV band 
  is $1.9\times 10^{40}$ erg s$^{-1}$ for the {\sc (bb+po)} model.  
For X10, the 0.3--10 keV isotropic luminosity 
  (from Table~\ref{tab:fits_comb_x10})  
  is $1.3\times 10^{40}$ erg s$^{-1}$.   
To estimate the total UV/X-ray luminosity of X7  
  we  integrated the flux of the thermal blackbody component down to 10~eV.  
For the power-law component, 
  we retained the 0.3$-$12 keV range from our observation,   
  to avoid introducing arbitrary cutoffs.  
This gives a conservative lower limit 
  to the contribution of the power-law component.    
The total UV/X-ray luminosity for X7 for the {\sc bb+po} model    
  is therefore $2.1\times 10^{40}$ erg s$^{-1}$,  
  of which $\sim 8\times 10^{39}$ erg s$^{-1}$ 
  is from the blackbody component. 
Using the {\sc diskbb+po} model 
  increases the luminosity to $\sim 3.2\times 10^{40}$ erg s$^{-1}$,  
  of which $\sim 1.8 \times 10^{40}$ erg s$^{-1}$ 
  is from the disk blackbody.   
For X10, we have only the power-law component,  
  and so the conservative lower limit 
  of the isotropic UV/X-ray luminosity 
  is $1.3\times10^{40}$ erg s$^{-1}$. 

These luminosities are uncertain 
  to the extent of the accuracy of the distance to NGC4559 
  and of the applicability of the spectral models. 
The Sanders et al.\ (2003) distance of 9.69~Mpc for NGC4559    
  is calculated from the redshift, 
  using the cosmic attractor model in Mould et al.\ (2000).   
The value is the same as that calculated by Tully (1988).  
However, the distance to NGC4559 has not been measured 
  with primary distance indicators.   
The intrinsic motion of the NGC4559 relative to the Hubble flow 
  may affect the distance derived from redshifts, 
  even when the correction has been made for our Galaxy. 
NGC4559 is a member of the Coma II cloud (Tully 1988), 
  for which the only Cepheid distance measurement is 
  that for NGC4725 at 12.7 Mpc (Gibson et al.\ 1999). 
The distance of 9.7 Mpc should therefore be considered 
  as a sensible lower limit.  
As noted above, the luminosity/flux for X7 that we obtained  
  depends on the assumed spectral models for the soft component.   
Nevertheless, all of the models yield luminosities 
  exceeding $10^{40}$ erg s$^{-1}$ (in the 0.3--10 keV band),  
  and in some epochs the total UV/X-ray luminosity 
  reaches $\sim 4\times10^{40}$ erg s$^{-1}$.  
The derived total UV/X-ray luminosities of X10 
  are approximately half those of X7, 
  from $8\times 10^{39}$ erg s$^{-1}$ 
  (for the first \chandra\/ observation)  
  to $1.5\times 10^{40}$ erg s$^{-1}$ 
  (for the second \chandra\/ observation).  
X7 and X10 are therefore among the most luminous ULX known.   
These luminosities also require very high mass accretion rates,  
  ${\dot M} \sim 1.7 \times 10^{-6} (0.1 /\eta)$ \Msun\/ yr$^{-1}$,  
  where $\eta (\equiv L / {\dot M} c^2)$ is the efficiency 
  of converting gravitational energy into radiation.   

The X-ray luminosities of X7 and X10  clearly exceed the Eddington limit 
  for 20 \Msun\/ compact objects with spherical accretion.   
This implies that both X7 and X10 could be IMBH
  if the emission is isotropic and the accretion is spherical. However, when these conditions are not satisfied, for example through inhomogeneity in the accreting material induced by instabilities  such as those discussed by Begelman (2002), it may not be necessary to invoke an IMBH to explain the observed luminosity.

\subsection{The nature of the compact accretor from the X-ray spectra}

\subsubsection{The blackbody component}

The spectra of X7 show a soft thermal blackbody component, 
  thus its bright X-ray luminosity cannot be easily explained 
  by models assuming relativistic beaming of synchrotron radiation.   
The effective temperature of the blackbody component 
  is $T_{\rm eff} \sim 0.12$ keV.  
By assuming a ``spherical'' geometry,  
  a {\sc bb} temperature of 0.12 keV and 
  a luminosity of $0.8\times 10^{40}$ erg s$^{-1}$
  (Table~\ref{tab:xmm_fits_x7}), together with a distance of 9.7 Mpc, 
  imply a characteristic radius of $1.8\times10^9$~cm 
  for the emission region.  

The nature of this characteristic radius is open to interpretation.   
If we assume that the thermal blackbody radiation originates  
  from a compact opaque electron scattering sphere around a black hole,  
  then the radius corresponds to the photospheric radius $R_{\rm ph}$ 
  of the scattering sphere. 
King \& Pounds (2003) considered a wind/outflow model  
  for such scattering region, 
  in which the characteristic size $R_{\rm ph}$ 
  (which is defined by the photosphere),  
  and the effective temperature $T_{\rm eff}$ 
  are expressed in terms of the mass outflow rate ${\dot M}_{\rm out}$ 
  and the BH mass \Mbh. 
By eliminating ${\dot M}_{\rm out}$ in their equations (17) and (18),
  and inserting the values for $R_{\rm ph}$ and $T_{\rm eff}$  
  derived from our observation,  
  we obtain \Mbh $\simeq 800$\Msun.   

We note that in this scenario, 
  the emission is heavily absorbed 
  because of the dense wind material surrounding the system, as seen in Cyg X-3 (Heindl et al., 2003).   
Moreover, we also expect such wind material will be illuminated 
  by the powerful X-rays from the accreting source, 
  but no evident nebulosity is seen in the optical {\it HST} image 
  (Figure~7) 
  or in the narrow-band images in Mirioni (2003) 
  and Roberts et al.\ (2002).   

Alternatively, the radius can be interpreted 
  as the inner radius ($R_{\rm in}$) of the an accretion disk 
  from which the soft thermal component originates. 
In this case, we may estimate the BH mass directly 
  from the parameters of the fitted spectral models.     
The {\sc diskbb} model,  
  gives  $R_{\rm in} (\cos i)^{1/2} = 1.2^{+1.0}_{-0.5} \times 10^9$~cm. 
If we assume that the accretion disk is truncated 
  at the innermost stable circular orbit, 
  then \Mbh$\sim 1.4^{+1.2}_{-0.6} \times 10^3 (\cos i)^{-1/2}$\Msun\/   
  for a Schwarzschild BH.   
From the {\sc bmc} model,  
  we obtain, instead, an effective radius $R_{\rm eff}$  
  and a colour tempertaure $T_{\rm col}$ 
  (Schrader \& Titarchuk 1999). 
By assuming a hardening factor $T_{\rm col}/T_{\rm eff} = 2.6$, 
  a value applicable for BH systems, such as GRO J1655$-$40,  
  $T_{\rm eff}$ together with $R_{\rm eff}$  
  implies a BH mass $\sim 1.7\pm 0.7 \times 10^3 (\cos i)^{-1/2}$\Msun 
  (see Schrader \& Titarchuk 2003 for details).  

We note that all these   arguments
  generally yield \Mbh\ $\sim10^3$\Msun\/ with $L \sim 0.1 L_{Edd}$. 
If we take these masses at face value, 
  X7 is a strong IMBH candidate.  

The characteristic radius 
  is not necessarily related to the last stable circular orbit 
  around a BH.  
The soft blackbody component may arise from a larger region.  
For an accretion disk truncated 
  at a radius much larger than the innermost stable circular orbit,      
  the same fitted temperature implies 
  a smaller \Mbh\/ but a larger  $L/L_{Edd}$ ratio.  
For example, 
  a system with \Mbh$\approx 200$\Msun, $L\simeq L_{Edd}$,  
  and an accretion disk truncated at an inner radius $R \sim 60 R_g$ (where $R_g=GM_{\bullet}/c^2$ is the gravitational radius of a BH)
  would also have explained the low characteristic temperature 
  deduced from the spectral fits.    
However, in this scenario 
  the emission from the region within the truncation radius 
  must be suppressed,  
  and it can be contrained observationally 
  by the flux of the Comptonised (power-law) component. 
In order for the emission from this region not to exceed  
  the value observed in the power-law component, 
  a large fraction, $>90$ percent, of the radiation must be advected.  
A very large $\dot M$ is thus required 
  to explain the observed X-ray luminosity. 
The situation becomes worse 
  when smaller \Mbh\/ (and higher $L/L_{Edd}$) are considered. 
For a system with \Mbh$\approx 20$\Msun\/ and an accretion disk truncated 
  at an inner radius $R \sim 600 R_g$,
  then $L \simeq 10L_{Edd}$. 
In this case, the advection is required to be extreme, 
  with the accreting matter radiating only $\sim 1$ percent of its gravitational energy. Whether this case is viable requires further theoretical investigation. 

\Mbh\/ may be also reduced if we allow the radiation to be anisotropic. 
A modest beaming factor $b \sim 0.1$, 
  reduces \Mbh\/ by a factor of 10. 
Only very extreme values of $b$   
  can reduce the black-hole mass from $10^3$ to $<20$\Msun.   However,  Madau (1988) calculated beaming factors from thick disk structures to be $<5$, and Strohmayer \& Mushotsky (2003) argued that the QPOs seen in the M82 X-1 source precluded substantial beaming.
This poses difficulties 
  for anistropic radiation (non-relativistic beaming) models,  
  such as that suggested in King et al. (2001). 

Our conclusion above rests mainly on the high luminosity 
  and low effective temperature for the thermal (blackbody) component 
  obtained in the spectral fits for X7.   
We are aware that the temperatures obtained 
  from the {\sc bb} and {\sc diskbb} models  
  may not be very reliable parameters to determine the BH mass.   
Ebisawa st al.\ (2003) found 
  that some ULX (e.g.\ ``Source 1'' in IC~342) can be explained  
  more self-consistently 
  within the conventional framework 
  of accretion onto stellar-mass ($\sim 10-20$ \Msun) BH,    
  when a slim-accretion disk model (see Wararai et al.\ 2000) 
  is used in the spectral fitting 
  instead of the models derived 
  from standard optically thick geometrically thin disks 
  (Shakura \& Sunyaev 1973).  
ULX such as Source 1 in IC~342 
  tend to have high fitted temperatures ($\sim 1$~keV),   
  for the thermal blackbody component.   
However, X7 in NGC4559 is probably different from these sources 
  since its thermal component has a much lower effective temperature. Similar cool temperatures are seen in thermal components of NGC1313 X-1 and X-2 (Miller et al 2003).

\subsubsection{The power-law components}   

The \xmm\/ spectra of X7 and X10 clearly show a power-law component.    
This eliminates the possibility of nuclear burning on the surface of a white dwarf, 
  as investigated, for example, by Fabbiano et al. (2003) 
  for the Antennae source CXOAnt J120151.6$-$185231.9.

By analogy with the spectral classification of Galactic BH systems, 
  X7 and X10 may be considered 
  in a soft and hard spectral state respectively. 
The conventional explanation for the power-law component 
  in the X-ray spectra of these systems   
  is Compton emission from hot electron clouds 
  (Sunyaev \& Titarchuk 1980; Pozdnyzkov, Sobol and Sunyaev 1983; 
  Zdziarski 1985). 
The accretion disk provides the seed photons,   
  which are up-scattered to the keV energy range 
  by the energetic electrons in a hot corona of an accretion disk  
  or in a hot cloud surrounding the BH event horizon.   
The spectral index of the power-law depends mainly on 
  the scattering optical depth  
  and the temperature of the energetic electrons 
  (for a thermal cloud). 
The photon index of the power-law component of X7    
  is $\Gamma \approx 2$, corresponding to spectral index $\approx 1$, 
  a value common to AGN and Galactic X-ray binaries.     
For modest scattering optical depths, say $\tau \approx 1$, 
  this spectral index implies effective temperatures 
  of about 40~keV and 85~keV for the electron clouds 
  with a slab and a spherical geometry respectively  
  (Titarchuk \& Lyubarskij 1995).      
The electron temperatures will be higher for smaller optical depths.  If we therefore extrapolate the luminosities of the power-law components to up to 50~keV, the total UV/X-ray luminosities for X7 and X10 rise to $\simeq 3.5\times10^{40}$ (for the {\sc (diskbb+po)} fit) and $\simeq 1.7\times10^{40}$ (for the {\sc po}) fit respectively.

Alternatively, 
  the disk photons can be up-scattered by non-thermal electrons,   
  such as those in bulk relativistic motion 
  near the BH event horizon  
  (Titarchuk, Mastichiadis \& Kylafis 1996) 
  or in relativistic jets 
  (Georganopoulos et al. 2002; K\"{o}rding, Falcke \& Markoff 2002).    
Photons Compton up-scatted by these non-thermal electrons 
  also have power-law spectra.    
It can be shown that Comptonised emission from relativistic jets 
  from stellar-mass BH can be Doppler-boosted 
  and amplified by factors $> 10^3$, 
  making the luminosity appearing super-Eddington 
  (Georganopoulos et al. 2002). 
However, such amplification can be observed only in a narrow viewing cone 
  in which the jets are along the line of sight. Moreover, the observed luminosities are at the upper limit of what this process can produce for systems with parameters similar to Cyg X-1.  It may be suitable for X10, but not X7, with its high luminosity soft component.   

It has been suggested that 
  the power-law X-ray component from BH binaries 
  in the (low)-hard spectral state 
  is due to direct synchrotron radiation (Fender 2001).   
If the observed X-ray spectrum from X10 is due to synchrotron radiation, the implied 
Lorentz factor $\gamma > 10^5$,  
  and substantial power is also emitted in the optical and radio bands 
  by an even larger number of electrons with lower $\gamma$.   
X10 is not listed in the FIRST radio catalogue (Becker, White \& Helfand, 1995) (nor is X7),  
  and its optical counterpart is not visible 
  in the {\it HST} image of the field (Figure~7) at the level of $R$ \ltae\/ 23.5 and $B$ \ltae\/ 24.5. 
Moreover, the large-amplitude variability on rapid timescales 
   expected in the relativistic beaming synchrotron model 
  for ULX (see Ulrich, Maraschi \& Urry 1997)
  is not seen in the data of X10 (Figure~\ref{fig:xmm_pds}).  
The power-law component is therefore 
  unlikely to be direct synchrotron emission.   
  
In view of the similar properties of the power-laws of X10 and X7, 
  their origins are probably the same, 
  {\it viz.} they are both due to Comptonised emission 
  from lower energy photons from the accretion disk. 
In sources such as X10,
  evidence of the soft thermal emission from the accretion disk can be hidden,  
  if it emanates from radii somewhat larger than the innermost stable orbit 
  and hence is softer. 
The soft disk emission is then overwhelmed 
  by the harder Comptonised radiation 
  from the hot electron clouds in the central region around the BH.  
If we assume that the accretion rate of X10 is sub-Eddington, 
  the luminosity of the power-law component requires \Mbh\/$> 100$~\Msun.    
However, for super-Eddington accretion, 
  a lower mass BH is permitted for the same observed luminosity. 
For \Mbh $< 20$~\Msun, the accretion rate 
  must exceed 5 times the Eddington rate.      

To explore the maximum luminosity of a {\sc bb} component consistent with the data, we included a  thermal {\sc bb} component with a fixed temperature of 0.12 keV . The best fit is obtained for no {\sc bb} component, with a 90 percent  confidence limit for the {\sc bb} normalisation of $< 1.76\times 10^{-6}$. The emitted flux from a thermal component in the 0.3--12 keV band is $< 1.0\times 10^{-13}$ erg cm$^{-2}$ s$^{-1}$, $< 8$ percent of the total emitted flux in this band. 
The total emitted X-ray/UV flux in the {\sc bb} component (0.01--12 keV) is $< 1.5\times 10^{-13}$ erg cm$^{-2}$ s$^{-1}$, corresponding to a  luminosity $< 1.7\times 10^{39}$ erg s$^{-1}$.
Consequently, if there is a blackbody-like component in X10 with a temperature of 0.12 keV then 
it is an order of magnitude weaker than that in X7. Lower temperature thermal components would also be consistent with the data, as they fall outside of the \xmm\/ band. This would imply higher masses for the compact object, if the component is interpreted as originating in an opaque scattering photosphere around it.

\subsubsection{Spectral residuals}

The spectral fit for X10 has residuals in the region of 2.9 keV. These deviations are evident in the counts spectrum in Figure~\ref{fig:all_spec_x10}.  Closer inspection shows that the residuals are in both \chandra\/ and \xmm\/ spectra, but their nature is different: in the first \chandra\/ observation there is a broad enhancement at this energy; by the time of the second \chandra\/ observation, there is an absorption cut into the enhancement -- the wings are still in emission, and by the time of the \xmm\/ observation there is only a narrow absorption feature.

The origin of this feature is unclear. It may be instrumental but we found no record of features at such energies in other \chandra\/ and \xmm\/ observations.  We therefore proceed on the assumption that it is real.

There are no strong spectral lines expected in this region, the only candidates being S {\sc xvi} at 2.62 keV and a fainter S {\sc xv} line at 2.88 keV. Such lines are not expected {\it a priori} and are not observed in other bright BH candidates.

We have considered whether X10 may be a background quasar. Supporting evidence would be the power-law spectrum and the absence of any break in the power spectrum (unlike that seen in X7). In this case the feature at 2.9 keV might be a gravitationally-redshifted Fe K$\alpha$ line. With $z = 1.2$,
the luminosity in the 2-10 keV range (using the {\sl ASCA} band for comparison) is $\sim1.1\times10^{46}$ erg s$^{-1}$
for the \xmm\/ observation, and $\sim 6\times10^{45}$ erg s$^{-1}$ for the first Chandra observation when the line is strongest. QSOs with this luminosity are bright
but not uncommon at $1 < z < 2$. Variability by a factor of 2 is also
normal. The fact that the emission line becomes weaker or disappears
as the QSO becomes brighter is also common (Baldwin 1977). However, we believe this explanation is unlikely, since at this X-ray flux level the typical optical brightness of a QSO is $R\sim17-19$, well above the upper limit of sources in the \hst\/ images. 
From the fitted values of $n_{\rm H}$ (Tables~\ref{tab:xmm_fits_x10} and \ref{tab:fits_comb_x10}) we may expect an intrinisic extinction $A_V\simeq2$ magnitudes corresponding to an extinction $\simeq4$ mag at $\lambda\simeq2700$ \AA\/ ($R$ band in the observer's frame). Even if we take this into account we would still expect an optical counterpart brighter than $R\simeq23$ (Comastri, 2003), which is not seen in the \hst\/ images. In addition, the space density of background sources at this flux level is 0.2 per  degree$^{2}$ (Rosati et al 2002) so the probability of being projected against the nuclear regions of NGC4559 is small, $\sim 5\times 10^{-4}$. Finally, luminous QSOs do not generally exhibit X-ray emission or absorption features (Nandra et al. 2000).

An alternative is that the feature is a Doppler-shifted line  
  intrinsic to the source.   
If it is a red-shifted line, the only potential candidate 
  is the Fe K$\alpha$ lines, and then $v/c = 0.66$. 
Redshifted lines can originate from infall or a receding jet. 
However, either of these would have difficulty  
  in producing the narrow absorption features 
  in the \chandra\/ 2 and \xmm\ data, 
  when the central source is expected 
  to dominate the continuum emission. 
If it is a blue-shifted line, 
  there are a number of possibilities of lines 
  in the 1$-$2 keV range.  
Blue-shifting can be produced by outflow or an approaching jet. 
It is puzzling that only one line is seen. 
This may imply an abnormal abundance of particular elements 
  or a particular ionisation/radiation process.  
However, the {\sc tbvarabs} fits to the \xmm\/ data 
  do not indicate any abundance anomalies.  
The progressive change of the feature from emission to absorption, 
  evident in Figure~\ref{fig:all_spec_x10}, 
  could be explained in term of the jet/outflow scenario  
  by condensation and cooling of the outflowing material 
  as its distance from the central X-ray source increases.    
  
In X10, the upper limit on the equivalent width for a narrow Fe K line at 6.4 keV is  80 eV, corresponding to a flux of $4\times 10^{-7}$ ph cm$^{-2}$ s$^{-1}$. The upper limit on a broader line depends on the fitted line model: for a gaussian line of width $\sigma= 0.5$ keV, the upper limit on the equivalent width increases to 200 eV. In X7 the upper limits on the equivalent width for narrow and $\sigma=0.5$ keV Fe K lines at 6.4 keV is 130 eV and 45 eV  respectively. It is not clear what the absence of an Fe K line means. In the Galactic BH the line strengths are variable and upper limits on the equivalent width of this order are frequently obtained. As an example, in the case of Cyg X-1, the upper limit on a narrow line from ASCA is $\simeq30$ eV  (Ebisawa et al. 1996). 

\subsection{The Nature of the compact component from the timing analysis}

Based on the similarity of the power spectrum in Galactic BH binaries and in AGN, many studies have investigated the possible link between the characteristics of the power spectrum and the AGN mass (for recent examples see Markowitz et al. 2003, Uttley, McHardy \& Papadakis 2002, Czerny et al. 2001). The power spectra are described by broken power-laws with characteristic break frequencies which appear to be related to the mass of the compact object. This is reinforced by the extrapolation to the Galactic BH domain, where a scaled break frequency of $\sim$Hz is compatible with \Mbh$\sim10$\Msun, derived using binary kinematic arguments in Cyg X-1 (Hasinger \& Belloni 1990, Nowak et al. 1999a). 

The calibration of the  \Mbh$-f_b$ relationship is still relatively insecure. At lower masses, the relationship rests mainly on that for Cyg X-1 (although breaks at $\sim$Hz are also seen in other BH candidates), while for the higher mass end, estimates of AGN masses are based on a relatively small sample determined mainly from reverberation studies. The physical connection between the two scales is also based on general timescale arguments (models include bright knots in the accretion disk -- Lehto 1989, and Comptonisation -- Kazanas, Hua \& Titarchuk 1997), but more work is required to examine this in detail. For example, the cause for the observed range in power-law slopes is not clear, and other Galactic BH binaries such as GX339--4 show more complex power spectra, which have been modelled with QPOs (Nowak, Wilms \& Dove 1999b). Nevertheless, for several AGN there are consistency checks (for example Edelson \& Nandra 1999, Shih, Iwasawa \& Fabian 2003), and the scaling of variability timescales is considered to be broadly valid. If this is accepted, then it is possible to use the power spectra in Figure~\ref{fig:xmm_pds} to infer information on the masses of X7 and X10.

The situation is, however, complicated in that the power spectrum of Cyg X-1 in the low-hard state is seen to have two breaks (Hasinger \& Belloni 1990). One, $f_{b2}$, occurs at a few Hz, where the slope flattens from $\alpha\sim-2$ to $\sim-1$, and a second, $f_{b1}$, occurs at a 0.04-0.4  Hz (depending on the accretion rate, Hasinger \& Belloni 1990) where the slope is consistent with being flat $\alpha\sim0$. In the soft-high state there appears to be no $f_{b1}$ (Reig, Papadakis \& Kylafis 2002), and the slope continues as $\alpha\sim-1$ to low frequencies. Some AGN have also been observed with double cutoffs (for example see Markowitz et al. 2003 and Uttley et al. 2002). The  \Mbh$-f_b$ relationship depends on which of the cutoffs is used at both high-and low-mass ends. Based on the index changes, Markowitz et al. (2003) suggest that the breaks in their sample correspond to the high frequency $f_{b2}$ breaks. In this case \Mbh$=37/f_b$\Msun.  (Markowitz  et al. (2003) excluded Akn 564 from their fits as they considered the break in this source at $\sim7.3$ day to be a low frequency $f_{b1}$ break. However, it
fits relatively well within their \Mbh$=37/f_b$\Msun\/ scaling relationship, yielding a
predicted mass of $2\times10^7$\Msun, which is within a factor of $\sim2$ 
of the reverberation mapping mass of $0.8\times10^7$\Msun.) If on the other hand, these are considered to be the low frequency $f_{b1}$ breaks, as in the fits to the knee model in Uttley et al. (2002), then the relationship is \Mbh$=1/f_b$ \Msun. 

We find for X7 (Figure~\ref{fig:xmm_pds}) that $f_b=0.028$ Hz. The formal fits to the slope $\alpha_1=0.20 \pm 0.14$ (Section~\ref{sec:obs_variability}) suggest that this cutoff is the lower frequency one $f_{b1}$. On the other hand, the product $P(f_b) \times f_b \sim 0.02$, which is
comparable to the same product, $\sim0.01$, measured by Markowitz et al. (2003),
{\it between} the lower and upper break frequencies. It appears not to be possible to unambiguously select which cutoff to adopt. The indicated masses are \Mbh$=38$\Msun\/ if $f_b=f_{b1}$ and 1300\Msun\/ if $f_b=f_{b2}$. If the break is instead to be interpreted as a QPO, then we note that the frequency is a factor $\sim10$ lower than the ``low frequency'' QPOs seen in the stellar-mass BH candidates GX339--4 (Rieg et al. 2002) and RXTE J1118+480 (Belloni, Psaltis \& van der Klis 2002).

It is evident from Figure~\ref{fig:xmm_pds} that any break in the power spectra of  X10 is at frequencies $<10^{-4}$ Hz. The scaling laws  above would imply masses of \Mbh$\sim10^4$ and $10^{5}$\Msun\/ in the two cases. 
 
\subsection{The ULX environment}

Fabbiano \& White (2003) discuss the environments of ULX in some detail. They note that in some cases ULX are associated with star forming regions, while in others, such as the Antennae, the ULX are offset from any star forming regions (Zezas \& Fabbiano 2002). In some cases, such as NGC5204 X1 and M81 X1, an optical counterpart to the ULX has been identified (Roberts et al. 2002, Pakull \& Mirioni 2002, for the former, Liu, Bregman \& Seitzer 2002 for the latter): there the counterparts appear to be early-type stars.

Outr \hst\/ images permit a detailed inspection of the environments of X7 and X10.
For X7 there are five candidates within the error circle, of which the faintest and bluest is visible only in the F450W filter. One other similarly faint and blue candidate is on the edge of the error circle. Of course it is not necessarily the case that any of these stars is the true counterpart to the ULX secondary, since this may be, for example, a giant (which would not be visible) rather than supergiant. For X10 there are no obvious optical counterparts in the error circle. However, this may be partly accounted for by the shorter exposure and coarser sampling of the X10 field in the \hst\/ observation. A detailed examination of the environment of the two ULX and any effects of reprocessing of the X-ray luminosity will be presented in a companion paper (Soria et al, in prep).
  
As noted in Section~\ref{sec:intro}, IMBH can be produced in low metallicity environments (Heger et al 2003), such as may be the case at least for X7, or as a primordial population in the first burst of star formation (Madau \& Rees 2001). Volontieri, Haardt \& Madau (2003) have examined the subsequent histories of these objects as they are incorporated in larger galaxies through hierarchical merging. Many will be incorporated in the central supermassive BH, but others remain in the halos or are ejected by three- or more body interactions (see Valtonen et al. 1994). They estimate that for a galaxy the size of the Milky Way (such as NGC4559), there will be $\sim10$ wandering IMBH with \Mbh\/ $150-1000$\Msun. The presence of two sources of the nature of X7 and X10 is therefore not surprising in this scenario. The question then is the expected frequency of capture of companion objects to fuel the accretion luminosity. These could be existing stars, or, more likely, new stars created by the IMBH triggering new star formation. This may be the reason for the association of some IMBH with star forming regions (Fabbiano et al. 2001), although this association is not entirely clear (Fabbiano \& White 2003).

Another channel discussed in Section ~\ref{sec:intro} for the production of IMBH, this time non-primordial, may be through the mergers of stellar mass BH produced in star clusters or super-star clusters. The cluster will harden through interactions (Lee 1995), especially as it is perturbed by the environment, which may also be changing as a result of any hierarchical merging taking place. Such cluster remnants could also carry within them lower mass stars which may be the source of the material for accretion, when these are captured. This may circumvent some of the difficulties raised by King et al (2001). The motion of these remnants through newly available gas when galaxy mergers take place may also provide the gravitational perturbation for new star formation regions. The ULX may appear displaced from the centre of the star forming region, depending on the time delay between the perturbation and the development of the star cluster.

\section{Conclusions}
\label{sec:con}

In these investigations of the ULX in NGC4559 we have a useful conjunction of circumstances: observations from three great observatories of two sources at the upper end of the luminosity range of ULX. The X-ray spectra obtained using \xmm\/ are among the highest signal/noise ratio achieved for such sources, with minimal nuclear contamination. These provide strong constraints on the spectral model fits. One of the objects, X7, shows both a blackbody and a power-law component in the X-rays, while the other, X10, has only a power-law component. The high count-rate permits an investigation of the temporal properties of the source to short timescales. The X-ray imaging from \chandra\/ permits an accurate localisation of the sources within NGC4559, and  \hst\/ provides deep imaging in the vicinity of the X-ray sources.

The nature of ULX sources has been vigorously debated (Section~\ref{sec:intro}). Our data strongly constrain the interpretation of these particular sources. We can immediately exclude  supernova remnants, or nuclear burning on a white dwarf surface. We have interpreted our data in terms of  accreting BH.

In the case of X7 there are two lines of argument on which we have based our conclusions: the luminosity of the blackbody-type component, and the temporal properties of the source. In the first line of argument, the fluxes are derived directly from the spectral fits. The UV/X-ray luminosities of this component can be inferred to be $0.8-1.8\times10^{40}$ erg s$^{-1}$. It is difficult to construct a scenario where significant beaming $(b<0.1)$ of this component can be achieved even if the emitting material itself is collimated. An IMBH accreting at $L\sim0.1L_{Edd}$ provides a simple explanation for the observed luminosities and blackbody temperatures, and at the same time a consistent explanation for the power-law component in terms of Comptonised emission. The masses derived from the disk fitting, or simple blackbody area measurements indicate that \Mbh$\sim2\times10^3$\Msun. This could be modified, however, by the (modest) beaming and inclination factors.  On the other hand, a model for a lower mass BH (\Mbh$<20$\Msun) fed from an accretion disk truncated at a large radius cannot be ruled out, if most of the energy within this radius is advected with high efficiency ($\sim 99$ percent).   

In the second line of argument, the power spectrum exhibits a break at 0.028 mHz. If the relationship between break frequency and compact object mass can be interpolated between stellar and supermassive BH, then this indicates X7 contains an IMBH. The masses determined from the temporal characteristics are \Mbh$\sim40$\Msun\/ or \Mbh$\sim1.3\times10^3$\Msun, depending on the calibration used.

In the case of X10, the absence of the blackbody-type component prevents any inference of the mass from disk parameters. The BH mass can be determined only from the luminosity: if this is sub-Eddington then we infer an IMBH for the compact object, while if we allow super-Eddington accretion or relativistic beaming it can be $<20$\Msun. The power-law component has the same slope as that for X7, and we interpret this to arise by a similar mechanism, Compton scattering. Among the beaming models, Doppler-boosted Compton emission from a relativistic jet can explain the high observed fluxes without invoking an IMBH. The power-law is unlikely to be produced by direct  beamed synchrotron emission as there is no evidence of sufficiently bright optical and radio counterparts, and the large amplitude temporal signatures expected are absent. 

\section{ACKNOWLEDGMENTS}

We thank Lev Titarchuk for discussions.

This paper is based on observations obtained with \xmm\/, an ESA science mission with instruments and contributions directly funded by ESA Member States and the USA (NASA),  the NASA \chandra\/ Observatory and the NASA/ESA Hubble Space Telescope.

\end{document}